\documentclass[preprint,showpacs,preprintnumbers,amsmath,amssymb]{revtex4}

\usepackage{graphicx}
\usepackage{dcolumn}
\usepackage{bm}
\usepackage{hyperref}
\usepackage{array}
\hypersetup{colorlinks=false}

\begin{document}
\newcommand{\RR}[1]{[#1]}
\newcommand{\intsum}{\sum \kern -15pt \int}
\newfont{\Yfont}{cmti10 scaled 2074}
\newcommand{\Y}{\hbox{{\Yfont y}\phantom.}}
\def\O{{\cal O}}
\newcommand{\bra}[1]{\left< #1 \right| }
\newcommand{\braa}[1]{\left. \left< #1 \right| \right| }
\def\Bra#1#2{{\mbox{\vphantom{$\left< #2 \right|$}}}_{#1}
\kern -2.5pt \left< #2 \right| }
\def\Braa#1#2{{\mbox{\vphantom{$\left< #2 \right|$}}}_{#1}
\kern -2.5pt \left. \left< #2 \right| \right| }
\newcommand{\ket}[1]{\left| #1 \right> }
\newcommand{\kett}[1]{\left| \left| #1 \right> \right.}
\newcommand{\scal}[2]{\left< #1 \left| \mbox{\vphantom{$\left< #1 #2 \right|$}}
\right. #2 \right> }
\def\Scal#1#2#3{{\mbox{\vphantom{$\left<#2#3\right|$}}}_{#1}
{\left< #2 \left| \mbox{\vphantom{$\left<#2#3\right|$}} \right. #3
\right> }}


\title{Parity-violating photon circular polarization in $nd\rightarrow$ $^3H\gamma$ with effective field theory at thermal energy}

\author{M. Moeini Arani}%
\email{m.moeini.a@khayam.ut.ac.ir (corresponding author)}
\author{S. Bayegan}
\email{bayegan@khayam.ut.ac.ir}

\affiliation{ Department of Physics, University of Tehran, P.O.Box
14395-547, Tehran, Iran
}%

\date{\today}

\begin{abstract}
We study the parity-violating effects in $nd\rightarrow$ $^3H\gamma$
process using pionless effective field theory (EFT($/\!\!\!\pi$)) at
thermal energy. For the weak $NN$ interaction the parity-violating
Lagrangian contains five independent low-energy coupling constants
(LECs). One can fix the coupling constants by comparison of the
calculated observables with a sufficient number of experimental
data. So, we need five observables to obtain LECs. One of the five
parity-violating observables is the photon circular polarization
($P_\gamma$) in $nd\rightarrow$ $^3H\gamma$ process. We calculate
$P_\gamma$ at the leading order in terms of LECs. We compare our
results with the previous $P_\gamma$ calculation based on the model
of Desplanques, Donoghue, and Holstein.
\end{abstract}

\pacs{11.10.-z; 21.45.+v; 21.10.Hw; 25.40.Lw; 21.30.Fe}
\keywords{pionless Effective Field Theory, few body system, Parity Violation, radiative capture}
\maketitle

\section{Introduction}\label{sec:introduction}

The Effective Field Theory (EFT) approach provides a systematic and
model-independent framework to analyse the hadronic parity-violating
(PV) effects in few-body system
\cite{shi-lin,schindler-springer-review}. We use the pionless
effective field theory (EFT($/\!\!\!\pi$)) for the PV process which
is at the energies well below the pion production threshold.

Recently, the leading-order (LO) PV nucleon-nucleon ($NN$)
Lagrangian has been expressed in a variety of bases
\cite{phillips-schindler-springer,girlanda}. The LO PV Lagrangian in
the partial-wave basis is written in terms of five independent
unknown coupling constants corresponding to five different S-P wave
combinations. This Lagrangian has been applied for neutron-neutron,
proton-proton and neutron-proton longitudinal asymmetries without
dibaryon fields in EFT($/\!\!\!\pi$)
\cite{phillips-schindler-springer}.

The calculation of $P_\gamma$ in $np\rightarrow$ $d\gamma$ at
threshold was introduced with the PV dibaryon-nucleon-nucleon
($dNN$) Lagrangian in EFT($/\!\!\!\pi$). The PV $dNN$ Lagrangian is
constructed in terms of three weak coupling constants. In fact,
parts of the PV Lagrangian in dibaryon formalism are given by this
Lagrangian structure. Result for $P_\gamma$ was obtained in terms of
two unknown weak $dNN$ coupling constants for the parity mixing in
the spin-singlet and spin-triplet states \cite{shin-ando-hyun}.

The photon asymmetry $A_\gamma$ and circular polarization $P_\gamma$
in $np\rightarrow$ $d\gamma$ process were also derived using the PV
$NN$ lagrangian based on the partial-wave basis with and without
dibaryon fields. EFT($/\!\!\!\pi$) with dibaryon fields has proven
to be convenient for few-nucleon systems at low energies
\cite{schindler-springer}.

In the three-body sector, a systematic analysis of the
parity-conserving
 (PC) $nd\rightarrow$
$^3H\gamma$ process at threshold was carried out using the
EFT($/\!\!\!\pi$)
\cite{sadeghi-bayegan-grieshammer,sadeghi-bayegan}. In the present
work, we concentrate on the PV $nd\rightarrow$ $^3H\gamma$
amplitude. We will make predictions for the PV observable in
$nd\rightarrow$ $^3H\gamma$ process with the PV Lagrangian based on
the five independent LECs in the dibaryon EFT($/\!\!\!\pi$).

The analysis of the PV three-nucleon interaction (3NI) and
calculation of neutron-proton and neutron-deuteron spin rotation
have been performed in the EFT($/\!\!\!\pi$) framework. The results
show that no PV 3NI enters at LO or next-to-leading order (NLO) in
the nucleon-deuteron ($Nd$) system
\cite{Grieshammer-schindler,G-S-S}. According to these results we
adopt the PV Lagrangian for the two-nucleon interaction in the
present calculation.

Theoretical calculations of the PV photon asymmetry and circular
polarization of the outgoing photon in $nd\rightarrow$ $^3H\gamma$
process were previously
 performed with the potential model. These analyses were based on Desplanques,
Donoghue, and Holstein (DDH) meson exchange potential. The
calculated results were $A_\gamma^t=0.81\times10^{-6}$ and
$A_\gamma^t=0.61\times10^{-6}$ for the two super-soft-core (SSC) and
Reid-soft-core (RSC) potentials, respectively. The DDH calculation
of $P_\gamma$ was also given for the two SSC and RSC potentials,
$P_\gamma=-1.39\times10^{-6}$ and $P_\gamma=-1.14\times10^{-6}$,
respectively \cite{desplanques-benayoun}. Recently, the PV effects
in $nd\rightarrow$ $^3H\gamma$ were calculated for DDH and EFT weak
interaction potentials \cite{Song-L-G}. We compare these $P_\gamma$
values with our results in sect.\ref{result}.

Finally, the experimental observation for the PV photon asymmetry in
$nd\rightarrow$ $^3H\gamma$ process was measured at ILL \cite{23 of
shin-ando-hyun} and the result was presented as
$A_\gamma^t=(4.2\pm3.8)\times10^{-6}$. The experimental data for
$P_\gamma$ in this process has not been reported to this point at
the thermal energy.

Our paper is organized as follows. In sect.~\ref{PC sector} we
present briefly the PC formalism for the $Nd$ scattering and
neutron-deuteron radiative capture. The PV $Nd$ scattering and PV
$nd\rightarrow$ $^3H\gamma$ processes are explained in sect.~\ref{PV
sector}. We obtain the PV amplitude of $nd\rightarrow$ $^3H\gamma$
process in sect.~\ref{PV nd radiative capture sector} with the
calculation of the appropriate diagrams. The numerical methods for
the evaluation of the PC and PV amplitudes of $nd\rightarrow$
$^3H\gamma$ process are explained in sect.~\ref{numerical}. In
sect.~\ref{result} the results of $P_\gamma$ in $nd\rightarrow$
$^3H\gamma$ are estimated and compared with the DDH framework. We
summarize the paper and discuss extension of the investigation to
other observable such as the PV photon asymmetry and other few-body
systems in sect.~\ref{conclusion}.

\section{Parity-conserving sector}\label{PC sector}

In the study of the weak interaction effects, we calculate the
observables which depend on the PV interactions. For this purpose
the $P_\gamma$ in $nd\rightarrow$$^3H\gamma$ process is chosen. To
this aim we need the PC amplitudes for both $Nd$ scattering and
$nd\rightarrow$$^3H\gamma$ processes. We briefly describe below the
procedure to obtain these two PC amplitudes at the leading order.
\subsection{Parity-conserving $Nd$ scattering}\label{PC nd scattering sector}
We proceed by introducing the Lagrangian of two- and three-nucleon
systems in Z-parametrization \cite{phillips-rupak-savage,20 of
sadeghi-bayegan}
\begin{eqnarray}\label{Eq:1}
  \mathcal{L}^{PC}=\mathcal{L}^{PC}_{two\:bady}+\mathcal{L}^{PC}_{three\:body}\,,
\end{eqnarray}
where
\begin{eqnarray}\label{Eq:2}
\mathcal{L}^{PC}_{two\;bady}=N^\dag\Big(i\partial_0+\frac{\nabla^2}{2m_N}\Big)N\qquad\qquad\qquad\qquad
  \nonumber\\ +d^{A^\dag}_{s}\Big[\Delta_s-c_{0s}\Big(i\partial_0+\frac{\nabla^2}{4m_N}+\frac{\gamma^2_s}{m_N}\Big)\Big]d^{A}_{s}
  \nonumber \\
  +d^{i^\dag}_{t}\Big[\Delta_t-c_{0t}\Big(i\partial_0+\frac{\nabla^2}{4m_N}+\frac{\gamma^2_t}{m_N}\Big)\Big]d^{i}_{t}\;\;\:
  \nonumber\\
  -y\Big(d^{A^\dag}_{s}(N^\dag P^A N)+h.c.\Big)\qquad\qquad\quad\;\;\;
  \nonumber \\ -y\Big(d^{i^\dag}_{t}(N^\dag P^i
  N)+h.c.\Big)+...\,,\qquad\qquad
\end{eqnarray}
\begin{eqnarray}\label{Eq:3}
  \mathcal{L}^{PC}_{three\; body}=\frac{m_N y^2 H_0(\Lambda)}{3\Lambda^2}\bigg(N^\dag\big(d^{i}_{t}\sigma_i\big)^\dag\big(d^{i}_{t}\sigma_i\big)N
  \nonumber\\ \qquad\qquad\qquad\qquad-\big[N^\dag\big(d^{i}_{t}\sigma_i\big)^\dag\big(d^{A}_{s}\tau_A\big)N+h.c.\big]\quad\;\;
  \nonumber \\+N^\dag\big(d^{A}_{s}\tau_A\big)^\dag\big(d^{A}_{s}\tau_A\big)N\bigg)+...\,,\quad\;\;
\end{eqnarray}
\begin{figure}
\includegraphics*[width=8.9cm]{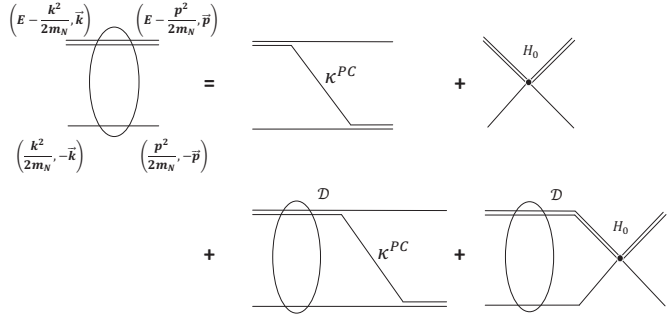}
\caption{\label{Fig:nd scattering}The Faddeev equation for the $Nd$
scattering. Single solid line denotes a nucleon. Double line is
propagator of the two intermediate auxiliary fields $d_s$ and $d_t$,
denoted by $\mathcal{D}$. $\mathcal{K}^{PC}$ and $H_0$ are the
propagator of exchanged nucleon and three-body interaction at LO,
respectively.}
\end{figure}
where $N$ is the nucleon iso-doublet and the auxiliary fields
$d^{A}_{s}$ and $d^{i}_{t}$ carry the quantum number of $^1S_0$
di-nucleon and the deuteron, respectively. The projectors $P^{i}$
and $P^{A}$ are defined by
$P^{i}=\frac{1}{\sqrt{8}}\sigma_2\sigma^i\tau_2$ and
$P^{A}=\frac{1}{\sqrt{8}}\sigma_2\tau_2\tau^A$ where $A=1,2,3$ and
$i=1,2,3$ are iso-triplet and vector indices, respectively.
$\tau^A$($\sigma^i$) are isospin (spin) pauli matrices. $m_N$ is the
nucleon mass and the three-nucleon interaction at the leading order
is $H_0(\Lambda)$ with cut-off $\Lambda$. The parameters
$\Delta_{s/t}$ are the mass differences between the dibaryons and
two nucleons.

According to the naive dimensional analysis: the parameters
$\Delta_{s/t}\sim Q$, with mass-dimension 1, are entered at LO;
dimensionless parameters $c_{0s/t}\sim Q^0$ first appear at NLO;
since they are expressed with two powers of momentum,
$c_{0s/t}p^2\sim Q^2$, where $Q$ is small momentum and appears in
$\frac{Q}{m_\pi}$ or $\frac{Q}{\tilde{\Lambda}}$ as an expansion
parameter ($m_\pi$ the pion mass, $\tilde{\Lambda}$ a symmetry
breaking scale) in EFT($/\!\!\!\pi$). The parameters of the
Lagrangians in eqs.(\ref{Eq:2}) and (\ref{Eq:3}) are fixed using
Z-parametrization \cite{20 of sadeghi-bayegan}. $d_{s/t}NN$ coupling
constant is chosen as
\begin{eqnarray}\label{Eq:0}
  y^2=\frac{4\pi}{m_N},
\end{eqnarray}
and the LO parameters $\Delta_{s/t}$ are obtained from the poles of
the $NN$ S-wave scattering amplitude at $i\gamma_{s/t}$.

The physical observables are cutoff independent, so in the
three-body systems at the leading order all-dependence on the cutoff
as $\Lambda\rightarrow\infty$ can be made to vanish by introducing
$H_0$. We know that the $H_0(\Lambda)$ is of order $Q^{-2}$
\cite{Bedaque-et.al,20 of sadeghi-bayegan}.

Neutron-deuteron scattering is shown schematically in
fig.\ref{Fig:nd scattering}. We denote the single straight line as a
nucleon with the propagator
$\frac{i}{q_0-\frac{q^2}{2m_N}+i\varepsilon}$. The double line
represents the dibaryon field with propagators for the singlet and
triplet cases. They are given by
\begin{eqnarray}\label{Eq:04}
  \mathcal{D}^{
  LO}(q_0,q)=\left(
                              \begin{array}{cc}
                                D^{
  LO}_{t}(q_0,q) & 0 \\
                                0 & D^{
  LO}_{s}(q_0,q) \\
                              \end{array}
                            \right),
\end{eqnarray}
where
\begin{eqnarray}\label{Eq:4}
  D^{
  LO}_{s(t)}(q_0,q)=\frac{1}{\gamma_{s(t)}-\sqrt{\frac{q^2}{4}-m_Nq_0-i\varepsilon}}\,,
\end{eqnarray}
where $\gamma_s=\frac{1}{a_s}$, $a_s$ is the scattering length in
$^1S_0$ state and $\gamma_t$ is the binding momentum of the
deuteron.

We work in the center of mass frame for the $Nd$ scattering with
$\vec{k}$ denotes the initial (on-shell) relative momentum of the
deuteron and the third nucleon, and $\vec{p}$ is the final
(off-shell) momentum. we use notation suggested by Griesshammer in
\cite{20 of sadeghi-bayegan}.

For scattering in the quartet ($S$=$\frac{3}{2}$) channel all spins
are aligned and there is no three-body interaction in this channel
because of the Pauli principle forbids the three nucleons to be at
the same point in space. In this channel, the initial and final
dibaryon fields have to be in the triplet channel. So, in this
channel we have only the $d_tN\rightarrow d_tN$ transition and the
corresponding amplitude in the cluster-configuration space can be
written as
\begin{eqnarray}\label{Eq:05}
    \left(
      \begin{array}{cc}
        t^{PC,(L)}_q(E;k,p) & 0 \\
        0 & 0 \\
      \end{array}
    \right)
    = -4\pi \mathcal{K}^{PC}_{(L)}(E,k,p)\left(
                               \begin{array}{cc}
                                 1 & 0 \\
                                 0 & 0 \\
                               \end{array}
                             \right)
    +\frac{2}{\pi}\int^\Lambda_0dq\:q^2\:\mathcal{K}^{PC}_{(L)}(E,q,p)
    \nonumber\\ \times\left(
                               \begin{array}{cc}
                                 1 & 0 \\
                                 0 & 0 \\
                               \end{array}
                             \right)
                              \mathcal{D}^{
  LO}(E-\frac{q^2}{2m_N},q)\left(
      \begin{array}{cc}
        t^{PC,(L)}_q(E;k,p) & 0 \\
        0 & 0 \\
      \end{array}
    \right).\:
\end{eqnarray}

For the $Nd$ scattering in the doublet ($S$=$\frac{1}{2}$) channel,
the parity-conserving amplitudes in cluster-configuration space are
given by
\begin{eqnarray}\label{Eq:5}
\left(
  \begin{array}{cc}
    t^{PC,(L)}_{d_{d_tN\rightarrow d_tN}}(E;k,p) & t^{PC,(L)}_{d_{d_sN\rightarrow d_tN}}(E;k,p) \\
    t^{PC,(L)}_{d_{d_tN\rightarrow d_sN}}(E;k,p) & t^{PC,(L)}_{d_{d_sN\rightarrow d_sN}}(E;k,p) \\
  \end{array}
\right) = \qquad\quad\qquad\qquad\qquad\qquad\qquad\qquad\qquad\qquad\nonumber\\
2\pi\left[\mathcal{K}^{PC}_{(L)}(E,k,p)\left(
                                                  \begin{array}{cc}
                                                    1 & -3 \\
                                                    -3 & 1 \\
                                                  \end{array}
                                                \right)
+\delta_{L0}\mathcal{H}(E,\Lambda)\left(
                 \begin{array}{cc}
                   1 & -1 \\
                   -1 & 1 \\
                 \end{array}
               \right)
                                                            \right]\qquad\qquad\qquad\qquad
                                                            \;\nonumber\\-\frac{1}{\pi}\int^\Lambda_0dqq^2\left[\mathcal{K}^{PC}_{(L)}(E,q,p)\left(
                                                                                                                         \begin{array}{cc}
                                                                                                                           1 & -3 \\
                                                                                                                           -3 & 1 \\
                                                                                                                         \end{array}
                                                                                                                       \right)+\delta_{L0}\mathcal{H}(E,\Lambda)\left(
                                                                                                                                               \begin{array}{cc}
                                                                                                                                                 1 & -1 \\
                                                                                                                                                 -1 & 1 \\
                                                                                                                                               \end{array}
                                                                                                                                             \right)\right]
                                                                                                                                            \qquad\qquad\;\nonumber\\ \times\mathcal{D}^{
  LO}(E-\frac{q^2}{2m_N},q)\left(
  \begin{array}{cc}
    t^{PC,(L)}_{d_{d_tN\rightarrow d_tN}}(E;k,q) & t^{PC,(L)}_{d_{d_sN\rightarrow d_tN}}(E;k,q) \\
    t^{PC,(L)}_{d_{d_tN\rightarrow d_sN}}(E;k,q) & t^{PC,(L)}_{d_{d_sN\rightarrow d_sN}}(E;k,q) \\
  \end{array}
\right),\qquad\qquad\;\:
\end{eqnarray}
where $t^{PC,(L)}_{d_{d_xN\rightarrow d_yN}}$ denotes the
$d_xN\rightarrow d_yN$ transition amplitude ($x,y=s$ or $t$) in the
doublet channel. The propagator of the exchanged nucleon,
$\mathcal{K}^{PC}_{(L)}$, is
\begin{eqnarray}\label{Eq:07}
\!\mathcal{K}^{PC}_{(L)}(E,k,p)=\frac{1}{2}\int^1_{-1}\!\!d
(cos\theta)\frac{P_L(cos\theta)}{\!k^2\!+p^2-ME+kp\:cos\theta},
\end{eqnarray}
where $P_L(x)$ denotes the $L$-th Legendre polynomial of the first
kind and $\theta$ indicates the angle between $\vec{k}$ and
$\vec{p}$ vectors. $\mathcal{H}=\frac{2H_0}{\Lambda^2}+...$
specifies the three-body force that renormalizes the amplitude and
it is introduced for the S-wave ($\delta_{L0}$).
$E=\frac{3\vec{k}^2}{4m_N}-\frac{\gamma^2_t}{m_N}$ is the total
non-relativistic energy.

\subsection{Parity-conserving $nd\rightarrow$ $^3H\gamma$ system}\label{PC nd radiative capture sector}
\begin{figure}
\includegraphics*[width=6.5cm]{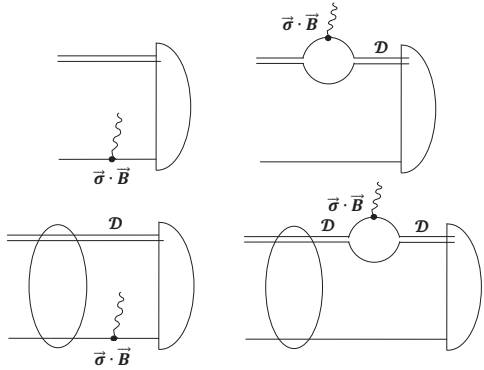}\centering
\caption{\label{Fig:PC nd capture}The leading-order diagrams which
contribute to PC (M1) amplitude of $nd\rightarrow$ $^3H\gamma$. The
small circles represent magnetic photon-Nucleon-Nucleon interaction.
Wavy line shows a photon. Dashed oval and dashed half-oval indicate
insertion of $Nd$ scattering amplitude at LO from fig.\ref{Fig:nd
scattering} and the formation of triton (the normalized triton wave
function). Remaining notation as fig.\ref{Fig:nd scattering}.}
\end{figure}

The dominated M1 amplitude of PC $nd\rightarrow$ $^3H\gamma$ process
receives the contribution from the neutron and proton magnetic
moment operators coupling to the magnetic field. At the leading
order, the magnetic M1 interaction of the photon with the single
nucleon is introduced by the Lagrangian
\begin{eqnarray}\label{Eq:6}
  \mathcal{L}_B=\frac{e}{2m_N}N^\dag\big(k_0+k_1\tau^3\big)\vec{\sigma}\cdot\vec{B}N\,,
\end{eqnarray}
where $k_0=\frac{1}{2}(k_p+k_n)=0.4399$ and
$k_1=\frac{1}{2}(k_p-k_n)=2.35294$ are the isoscalar and isovector
nucleon magnetic moments in the nuclear magnetons, respectively.
$k_p$ and $k_n$ denote the proton and neutron magnetic moments,
respectively. $e$ is the electric charge and $\vec{B}$ is magnetic
field. The diagrams of PC $nd\rightarrow$ $^3H\gamma$ at LO are
shown in fig.\ref{Fig:PC nd capture}. The dashed oval is the $Nd$
scattering calculated in sect.~\ref{PC nd scattering sector}. The
dashed half-oval is the normalized triton wave function.

We note that for obtaining the normalized triton wave function, one
should solve the homogeneous part of eq.(\ref{Eq:5}) with
application of $E=-B_t$ where $B_t$ is the binding energy of the
triton. The wave function is then normalized using the nucleon and
dibaryon propagators as briefly explained in \ref{Appendix A}. The
homogeneous part of eq.(\ref{Eq:5}) can be written as
\begin{eqnarray}\label{Eq:0005}
t^{^3H}(p)=
-\frac{1}{\pi}\int^\Lambda_0dqq^2\left[\mathcal{K}^{PC}_{(0)}(-B_t,q,p)\left(
                                                                                                                         \begin{array}{cc}
                                                                                                                           1 & -3 \\
                                                                                                                           -3 & 1 \\
                                                                                                                         \end{array}
                                                                                                                       \right)+\mathcal{H}(-B_t,\Lambda)\left(
                                                                                                                                               \begin{array}{cc}
                                                                                                                                                 1 & -1 \\
                                                                                                                                                 -1 & 1 \\
                                                                                                                                               \end{array}
                                                                                                                                             \right)\right]
                                                                                                                                            \nonumber\\ \times\mathcal{D}^{
  LO}(-B_t-\frac{q^2}{2m_N},q)\,t^{^3H}(q),\qquad\qquad\qquad\qquad\qquad\qquad\qquad\qquad\quad
\end{eqnarray}
where $t^{^3H}(q)=\bigg(\begin{array}{cc}
    t^{^3H}_{d_tN\rightarrow d_tN}(q) & t^{^3H}_{d_sN\rightarrow d_tN}(q) \\
    t^{^3H}_{d_tN\rightarrow d_sN}(q) & t^{^3H}_{d_sN\rightarrow d_sN}(q) \\
\end{array}\bigg)$. Generally, $t^{^3H}_{d_xN\rightarrow
d_yN}(q)$ denotes the contribution of the ${d_xN\rightarrow d_yN}$
transition ($x,y=s$ or $t$) for making the triton.

For PC sector at the lowest order, we have two dominated M1
transitions, to which $^3He$ and $^3H$ belong,
$j^P=\frac{1}{2}^+\rightarrow M1$ and $j^P=\frac{3}{2}^+\rightarrow
M1$. The following parametrization of the corresponding
contributions to the matrix element are,
\begin{eqnarray}\label{Eq:7}
  \big(t^\dag\sigma_aN\big)\big(\vec{\varepsilon}_d\times\big(\vec{\varepsilon}^\ast_\gamma\times\vec{\tilde{q}}\big)\big)_a\:,\qquad i\big(t^\dag
  N\big)\big(\vec{\varepsilon}_d\cdot\vec{\varepsilon}^\ast_\gamma\times\vec{\tilde{q}}\big)\,,\;\;\;
\end{eqnarray}
with $N$, $t$, $\vec{\varepsilon}_\gamma$, $\vec{\varepsilon}_d$ and
$\vec{\tilde{q}}$ are the 2-component spinors of the initial nucleon
field, the final $^3H$ (or $^3He$) field, the 3-vector polarization
of the produced photon, the 3-vector polarization of the deuteron
and the unit vector along the 3-momentum of the photons,
respectively. For both possible magnetic transitions with
$j^P=\frac{1}{2}^+ $ ($\mathcal{W}^{PC}(^2S_{\frac{1}{2}})$
amplitude) and $j^P=\frac{3}{2}^+$
($\mathcal{W}^{PC}(^4S_{\frac{3}{2}})$ amplitude) we can write
\begin{eqnarray}\label{Eq:007}
  \mathcal{W}^{PC}(^2S_{\frac{1}{2}})\big[t^\dag\big(i\vec{\varepsilon}_d\cdot\vec{\varepsilon}^\ast_\gamma\times\vec{\tilde{q}}+\vec{\sigma}\times\vec{\varepsilon}_d\cdot\vec{\varepsilon}^\ast_\gamma\times\vec{\tilde{q}}\big)N\big]\,,\!
\nonumber\\
\mathcal{W}^{PC}(^4S_{\frac{3}{2}})\big[t^\dag\big(i\vec{\varepsilon}_d\cdot\vec{\varepsilon}^\ast_\gamma\times\vec{\tilde{q}}+\vec{\sigma}\times\vec{\varepsilon}_d\cdot\vec{\varepsilon}^\ast_\gamma\times\vec{\tilde{q}}\big)N\big]\,.
\end{eqnarray}

The PC amplitude of $nd\rightarrow$ $^3H\gamma$ at the thermal
energy is calculated and then used in the evaluation of the
$P_\gamma$ in sect.\ref{result}.

\section{Parity-violating sector}\label{PV sector}
\subsection{Parity-violating $Nd$ scattering}\label{PV nd sector}
\begin{figure*}[tb]
\centering
\includegraphics*[width=17cm]{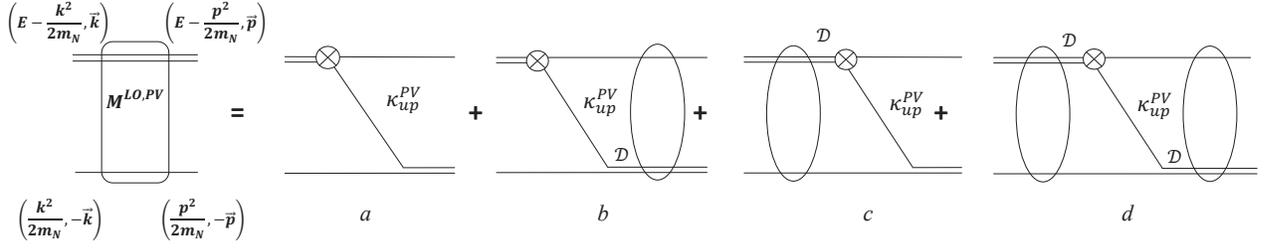}
\caption{\label{Fig2} The PV $Nd$ scattering diagrams at LO. The
dashed rectangular denotes the PV $Nd$ scattering amplitude. Circle
with a cross indicates the PV $dNN$ vertex. $\mathcal{K}^{PV}_{up}$
is the PV propagator of the exchanged nucleon when PV $dNN$ vertex
is in the upper side. Remaining notations are the same as
fig.\ref{Fig:nd scattering}. Time-reversed contributions not
displayed.}
\end{figure*}
The PV two-body transitions at the lowest order are
$^1S_0\leftrightarrow$ $ ^3P_0$, $^3S_1\leftrightarrow$ $^1P_1$ and
$^3S_1\leftrightarrow $ $^3P_1$. The leading-order PV Lagrangian
with dibaryon formalism is given by
\begin{eqnarray}\label{Eq:8}
   {\mathcal{L}_{PV}}=-\Big[g^{(^3S_1-^1P_1)}d^{i^\dag}_{t}N^T
   i\left(\overleftarrow{\nabla}\sigma_2\tau_2-\sigma_2\tau_2\overrightarrow{\nabla}\right)_iN\qquad\quad\qquad\qquad
 \nonumber \\
+g^{(^1S_0-^3P_0)}_{(\Delta I=0)}d^{A^\dag}_{s}N^T
i\left(\overleftarrow{\nabla}\sigma_2\sigma_i\tau_2\tau_A-\sigma_2\sigma_i\tau_2\tau_A\overrightarrow{\nabla}\right)_iN\qquad\;\,
\nonumber \\
+g^{(^1S_0-^3P_0)}_{(\Delta I=1)}\epsilon^{3AB}d^{A^\dag}_{s}\,N^T
i\left(\overleftarrow{\nabla}\sigma_2\sigma_i\tau_2\tau^B-\sigma_2\sigma_i\tau_2\tau^B\overrightarrow{\nabla}\right)_iN
\nonumber \\
+g^{(^1S_0-^3P_0)}_{(\Delta I=2)}\mathcal{I}^{AB}d^{A^\dag}_{s}N^T
i\left(\overleftarrow{\nabla}\sigma_2\sigma_i\tau_2\tau^B-\sigma_2\sigma_i\tau_2\tau^B\overrightarrow{\nabla}\right)_iN\;
\nonumber \\
+g^{(^3S_1-^3P_1)}\epsilon^{ijk}d^{i^\dag}_{t}N^T
\left(\overleftarrow{\nabla}\sigma_2\sigma^k\tau_2\tau_3-\sigma_2\sigma^k\tau_2\tau_3\overrightarrow{\nabla}\right)^jN\Big]\quad
\nonumber \\
+h.c.+...
\,.\qquad\qquad\qquad\qquad\qquad\qquad\qquad\qquad\quad\qquad
\end{eqnarray}
In the above equation, $g^{(\bar{X}-\bar{Y})}$ denotes the weak
$dNN$ coupling constant for the PV two-body transition between
$\bar{X}$ and $\bar{Y}$ partial waves. $\Delta I$ represents the
isospin change in the PV vertex and $\mathcal{I}$=diag(1,1,-2) is a
diagonal matrix in isospin space. The three-point in the last line
represents all transitions that will be neglected at the very low
energy.

For the three-nucleon systems, at the lowest order, there are four
possible transitions which mix S- and P-waves and conserve total
angular momentum,
\begin{eqnarray}\label{Eq:35}
  ^2S_{\frac{1}{2}}\leftrightarrow\,^2P_{\frac{1}{2}}\quad,\quad^2S_{\frac{1}{2}}\leftrightarrow
  \,^4P_{\frac{1}{2}}\;\:
  \nonumber \\ ^4S_{\frac{3}{2}}\leftrightarrow
  \,^2P_{\frac{3}{2}}\quad,\quad^4S_{\frac{3}{2}}\leftrightarrow
  \,^4P_{\frac{3}{2}}\,.
\end{eqnarray}

A systematic analysis by Griesshammer \textit{et al.}
\cite{Grieshammer-schindler} show that no parity-violating
three-nucleon interaction (PV 3NI) is participated in the
nucleon-deuteron system at the leading and next-to-leading orders in
EFT($/\!\!\!\pi$). So, in the three-body systems, since all
amplitudes converge at LO and NLO, no PV 3NI are needed for the
renormalization. Thus, we use the Lagrangian in eq.(\ref{Eq:8}) to
obtain the LO PV $Nd$ scattering amplitude for all possible channels
in eq.(\ref{Eq:35}).

The Feynman diagrams that contribute to the PV $Nd$ scattering
amplitude are shown in fig.\ref{Fig2} \cite{G-S-S}. In the
cluster-configuration space, the contribution of these diagrams in
fig.\ref{Fig2} can be written after doing integration over energy
and solid angle as
\begin{eqnarray}\label{Eq:035}
t^{LO,PV}_{up}(X\rightarrow Y;E,k,p)=4\pi
\mathcal{K}^{PV}_{up}(X\rightarrow
Y;E,k,p)\qquad\qquad\qquad\qquad\qquad\qquad
\nonumber\\-\frac{2}{\pi}\int^\Lambda_0dq
q^2\bigg[t^{LO,PC}(Y;E,q,p)\mathcal{D}^{LO}(E-\frac{q^2}{2m_N},q)\mathcal{K}^{PV}_{up}(X\rightarrow
Y;E,k,q)\quad
\nonumber\\
+\mathcal{K}^{PV}_{up}(X\rightarrow
Y;E,q,p)\mathcal{D}^{LO}(E-\frac{q^2}{2m_N},q)t^{LO,PC}(X;E,k,q)\bigg]+\quad
\nonumber \\
+\frac{1}{\pi^3}\int^\Lambda_0dq_1q^2_1\;\int^\Lambda_0dq_2q^2_2\;\bigg[t^{LO,PC}(Y;E,q_2,p)\mathcal{D}^{LO}(E-\frac{q^2_2}{2m_N},q_2)\qquad\qquad
\nonumber \\
\times \:\mathcal{K}^{PV}_{up}(X\rightarrow Y;E,q_1,q_2)\;\,
\mathcal{D}^{LO}(E-\frac{q_1^2}{2m_N},q_1)\;t^{LO,PC}(X;E,k,q_1)\bigg],
\end{eqnarray}
where $t^{LO,PC}(X or Y;E,k,p)$ is $2\times2$ matrix for LO PC $Nd$
scattering amplitudes, eqs.(\ref{Eq:05}), (\ref{Eq:5}), depending on
the incoming (outgoing) $X$ ($Y$) partial wave. $E=\frac{3k^2}{4
m_N}-\frac{\gamma^2_t}{m_N}$ and $\vec{k}$ $(\vec{p})$ are the
center-of-mass energy and the incoming (outgoing) momentum,
respectively. The PV kernel $\mathcal{K}^{PV}_{up}(X\rightarrow
Y;E,k,p)$ for all transitions with total angular momentum
$J=\frac{1}{2}$ is given by \cite{G-S-S} (see eq.(\ref{Eq:35}) for
all possible $X\rightarrow Y$)
\begin{eqnarray}\label{Eq:036}
\mathcal{K}^{PV}_{up}(^2S_{\frac{1}{2}}\rightarrow\,^2P_{\frac{1}{2}};E,k,p)=\frac{ym_N}{6\sqrt{2}\pi
kp}\,\big[2pQ_0(-\varepsilon)+kQ_1(-\varepsilon)\big]\left(
                                               \begin{array}{cc}
                                                 \mathcal{S}_1 & -\mathcal{T} \\
                                                 \mathcal{S}_1 & -\mathcal{T} \\
                                               \end{array}
                                             \right),
\nonumber \\
\nonumber \\
\mathcal{K}^{PV}_{up}(^2P_{\frac{1}{2}}\rightarrow\,^2S_{\frac{1}{2}};E,k,p)=\frac{ym_N}{6\sqrt{2}\pi
kp}\,\big[2pQ_1(-\varepsilon)+kQ_0(-\varepsilon)\big]\left(
                                               \begin{array}{cc}
                                                 \mathcal{S}_1 & -\mathcal{T} \\
                                                 \mathcal{S}_1 & -\mathcal{T} \\
                                               \end{array}
                                             \right),
\nonumber \\ \nonumber \\
\mathcal{K}^{PV}_{up}(^2S_{\frac{1}{2}}\rightarrow\,^4P_{\frac{1}{2}};E,k,p)=
\frac{ym_N}{3\pi
kp}\,\big[2pQ_0(-\varepsilon)+kQ_1(-\varepsilon)\big]\left(
                                               \begin{array}{cc}
                                                 \frac{\mathcal{S}_1-\mathcal{S}_2}{3} & \mathcal{T} \\
                                                 0 & 0 \\
                                               \end{array}
                                             \right),
\nonumber \\ \nonumber \\
\mathcal{K}^{PV}_{up}(^4P_{\frac{1}{2}}\rightarrow\,^2S_{\frac{1}{2}};E,k,p)=\frac{ym_N}{6\pi
kp}\,\big[2pQ_1(-\varepsilon)+kQ_0(-\varepsilon)\big]\left(
                                               \begin{array}{cc}
                                                 \mathcal{S}_2 & 0 \\
                                                 \mathcal{S}_2 & 0 \\
                                               \end{array}
                                             \right),\qquad\:
\end{eqnarray}
and with $J=\frac{3}{2}$ by
\begin{eqnarray}\label{Eq:0036}
\mathcal{K}^{PV}_{up}(^4S_{\frac{3}{2}}\rightarrow\,^2P_{\frac{3}{2}};E,k,p)=\frac{ym_N}{6\sqrt{2}\pi
kp}\,\big[2pQ_0(-\varepsilon)+kQ_1(-\varepsilon)\big]\left(
                                               \begin{array}{cc}
                                                 \mathcal{S}_2 & 0 \\
                                                 \mathcal{S}_2 & 0 \\
                                               \end{array}
                                             \right)Q^r_s,
\qquad\;\nonumber \\ \nonumber \\
\mathcal{K}^{PV}_{up}(^2P_{\frac{3}{2}}\rightarrow\,^4S_{\frac{3}{2}};E,k,p)=\frac{ym_N}{3\sqrt{2}\pi
kp}\,\big[2pQ_1(-\varepsilon)+kQ_0(-\varepsilon)\big]\left(
                                               \begin{array}{cc}
                                                 \frac{\mathcal{S}_1-\mathcal{S}_2}{3} & \mathcal{T} \\
                                                 0 & 0 \\
                                               \end{array}
                                             \right)Q^r_s,
\;\:\:\nonumber \\ \nonumber \\
\mathcal{K}^{PV}_{up}(^4S_{\frac{3}{2}}\rightarrow\,^4P_{\frac{3}{2}};E,k,p)=-\frac{\sqrt{10}ym_N}{6\pi
kp}\,\big[2pQ_0(-\varepsilon)+kQ_1(-\varepsilon)\big]\left(
                                               \begin{array}{cc}
                                                 \frac{\mathcal{S}_1-\mathcal{S}_2}{3} & 0 \\
                                                 0 & 0 \\
                                               \end{array}
                                             \right)Q^r_s,
\nonumber \\ \nonumber \\
\mathcal{K}^{PV}_{up}(^4P_{\frac{3}{2}}\rightarrow\,^4P_{\frac{3}{2}};E,k,p)=-\frac{\sqrt{10}ym_N}{6\pi
kp}\,\big[2pQ_1(-\varepsilon)+kQ_0(-\varepsilon)\big]\left(
                                               \begin{array}{cc}
                                                 \frac{\mathcal{S}_1-\mathcal{S}_2}{3} & 0 \\
                                                 0 & 0 \\
                                               \end{array}
                                             \right)Q^r_s,
\end{eqnarray}
where $\varepsilon=\frac{k^2+p^2-m_NE}{kp}$ and the
$(Q^r_s)^\alpha_\beta=\delta^r_s
\delta^\alpha_\beta-\frac{1}{3}(\sigma^r\sigma_s)^\alpha_\beta$ is
the spin quartet projector. The index $\alpha$ $(\beta)$ denotes the
spin of the outgoing (incoming) nucleon and $r$ $(s)$ is the spin
component of the outgoing (incoming) $^3S_1$ dibaryon. $Q_L(z)$ is
the $L$-th Legendre polynominals of the second kind with the complex
argument. The "up" subscript represents the PV vertex in the upper
side of the diagrams. The $\mathcal{S}_1$, $\mathcal{S}_2$ and
$\mathcal{T}$ are independent linear combinations of the PV coupling
constants, $g^{(\bar{X}-\bar{Y})}$,
\begin{eqnarray}\label{Eq:037}
\mathcal{S}_1=3g^{(^3S_1-^1P_1)}+2\tau_3g^{(^3S_1-^3P_1)},
\nonumber \\
\mathcal{S}_2=3g^{(^3S_1-^1P_1)}-\tau_3g^{(^3S_1-^3P_1)},
\:\;\nonumber\\
\mathcal{T}=3g^{(^1S_0-^3P_0)}_{(\Delta
I=0)}+2\tau_3g^{(^1S_0-^3P_0)}_{(\Delta I=1)},\:\;
\end{eqnarray}
where the value of the $\tau_3$ is +1 and -1 for the scattering on
the proton and neutron, respectively.

We note that the eq.(\ref{Eq:035}) is not a Faddeev equation.
However the Faddeev equations in the eqs.(\ref{Eq:05}), (\ref{Eq:5})
should be solved numerically for the PC $Nd$ scattering amplitude
and then replace the results in eq.(\ref{Eq:035}). We solve
eq.(\ref{Eq:035}) numerically.

The calculation of the time-reversed contributions of the diagrams
in fig.\ref{Fig2} is the same as explained above. We must only
substitute the $\mathcal{K}^{PV}_{down}(X\rightarrow Y;E,k,p)$ in
eq.(\ref{Eq:035}) with $\mathcal{K}^{PV}_{up}(X\rightarrow Y;E,k,p)$
to obtain the time-reversed contribution of the diagrams in
fig.\ref{Fig2}. The $\mathcal{K}^{PV}_{down}($ $X\rightarrow
Y;E,k,p)$ can be written as
\begin{eqnarray}\label{Eq:038}
\!\!\!\!\mathcal{K}^{PV}_{down}\!(X\rightarrow
Y;E,k,p)_{s\beta}^{r\alpha}\!\!=\!\!\Big[\mathcal{K}^{PV}_{up}\!(Y\rightarrow
X;E,p,k)^{s\beta}_{r\alpha}\Big]^\dag\!\!.
\end{eqnarray}

Finally, after multiplication of the wave function normalization
factor the physical amplitude of the neutron-deuteron scattering in
the on-shell point ($k=p$) is
\begin{eqnarray}\label{Eq:039}
T^{LO,PV}(X\rightarrow Y;E,k,k)=\left(
           \begin{array}{cc}
             \sqrt{\mathcal{Z}_{LO}} & 0 \\
           \end{array}
         \right)
t^{LO,PV}(X\rightarrow Y;E,k,k)\left(
                                 \begin{array}{c}
                                   \sqrt{\mathcal{Z}_{LO}} \\
                                   0 \\
                                 \end{array}
                               \right),
\end{eqnarray}
where $t^{LO,PV}$ is the sum of both $t^{LO,PV}_{up}$ and
$t^{LO,PV}_{down}$ and $\mathcal{Z}_{LO}=\frac{2\gamma_t}{m_N}$ is
the LO deuteron wave function normalization factor in
Z-parametrization.

\subsection{Parity-violating $nd\rightarrow$ $^3H\gamma$
system}\label{PV nd radiative capture sector}In this section, we
concentrate on the PV amplitude of $nd\rightarrow$ $^3H\gamma$
process. For this purpose, we need to obtain the electromagnetic
(EM) matrix elements that contribute at the thermal energy
($2.5\times10^{-8} \textrm{MeV}$) near threshold. In the low-energy
regime, we work with the Lagrangian introduced in eq.(\ref{Eq:8})
for the PV interaction.

At threshold, E1 transition contributes a dominate part in the PV
neutron-deuteron radiative capture process. Obviously, as we move
above the threshold we must also add the parity-violating M1
transition to the leading E1 amplitude.
\begin{figure*}[tb]\centering
\includegraphics*[width=16cm]{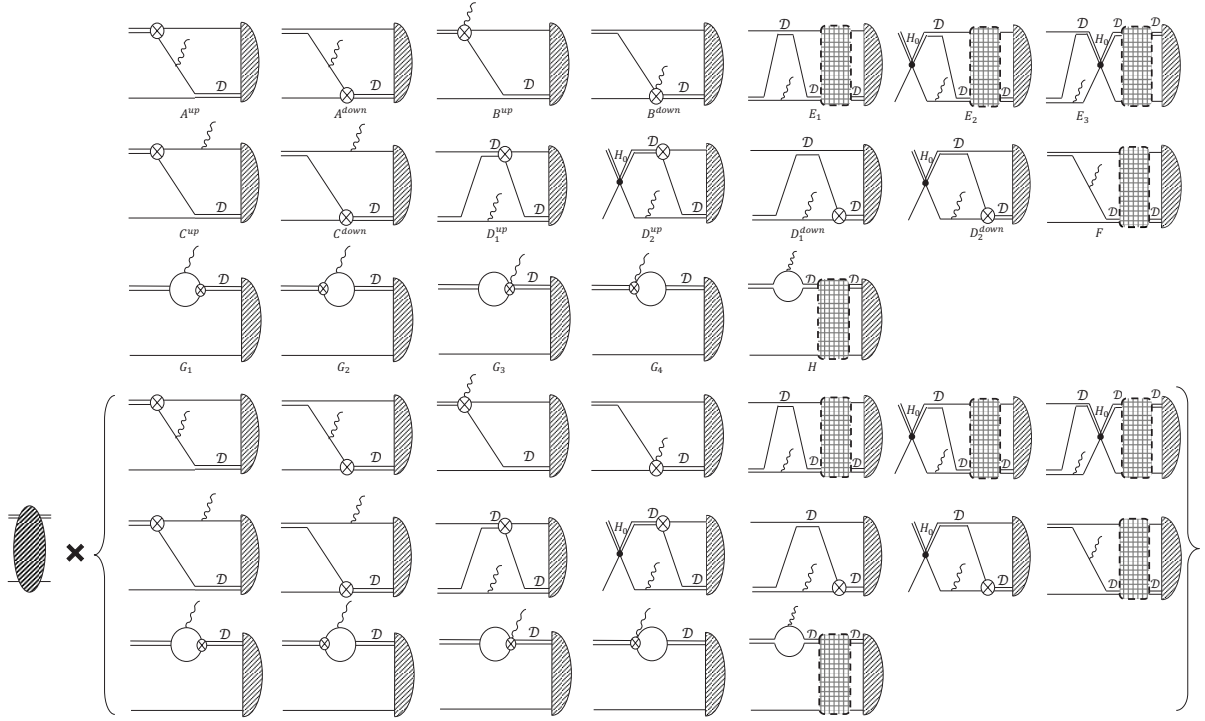} \caption{\label{Fig3}
The set (I) diagrams of the PV $nd\rightarrow$ $^3H\gamma$ process
at LO. Circle with a cross and wavy line is the PV $\gamma dNN$
vertex. The dashed rectangular with dashed line around it denotes
the PV $Nd$ scattering amplitude depicted in fig.\ref{Fig2} without
participation of the diagrams which have the full PC $Nd$ scattering
in the right hand side. $H_0$ is the three-body force which has been
introduced in fig.\ref{Fig:nd scattering}. All notations are the
same as the previous figures.}
\end{figure*}
\begin{figure*}[tb]\centering
\includegraphics*[width=10cm]{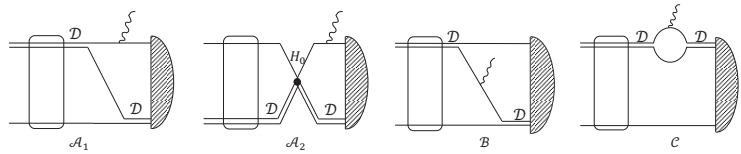} \caption{\label{Fig4}
The set (II) diagrams of the PV $nd\rightarrow$ $^3H\gamma$ process
at LO. The dashed rectangular with solid line around it is the PV
$Nd$ scattering amplitude shown in fig.\ref{Fig2}. All notation are
the same as the previous figures.}
\end{figure*}
In the present calculations, we introduce the diagrams that
contribute to the E1 PV neutron-deuteron radiative capture process.
These diagrams are shown schematically in figs.\ref{Fig3} and
\ref{Fig4}. We identify the diagrams in fig.\ref{Fig3} as the set
(I) and those of fig.\ref{Fig4} as the set (II). The dashed
rectangular with dashed line around it in fig.\ref{Fig3} denotes the
PV $Nd$ scattering amplitude corresponding to the diagrams $a$, $c$
and time-reversed contribution of $a$, $b$ shown in fig.\ref{Fig2}.
The dashed rectangular with solid line around it in fig.\ref{Fig4}
is the PV $Nd$ scattering amplitude depicted in fig.\ref{Fig2}. In
figs.\ref{Fig3} and \ref{Fig4} dashed oval and dashed half-oval are
the same as before.

We consider only the one-body current for E1 interaction and we
employ the "convection" nucleon current for the
photon-nucleon-nucleon ($\gamma NN$) vertex that can be written
generally by \cite{book}
\begin{eqnarray}\label{Eq:19}
J_N(\vec{r})=\frac{1}{2}\sum_j(1+\tau_3^{(j)})
\Big[\frac{\vec{P}}{2m_N}\:\delta(\vec{r}-\vec{r_j})
+\:\delta(\vec{r}-\vec{r_j})\:\frac{\vec{P^\prime}}{2m_N}\Big]\cdot\:\vec{\varepsilon_\gamma^\ast}\,,
\end{eqnarray}
where $\vec{P}$ and $\vec{P^\prime}$ are momenta for the incoming
and outgoing nucleons. $\tau^{(j)}_3$ and $\vec{r_j}$ are the
$3^{rd}$-component of the isospin operator and the position of the
$j^{th}$-nucleon, respectively. For the single nucleon, it is
reduced to
\begin{eqnarray}\label{Eq:20}
\frac{1}{2m_N}\:(1+\tau_3)\:\frac{1}{2}\:(\vec{P}+\vec{P^\prime})\cdot\vec{\varepsilon_\gamma^\ast}.
\end{eqnarray}
The PV photon-dibaryon-nucleon-nucleon ($\gamma dNN$) vertex in
fig.\ref{Fig3} is given by the minimal substitution of
$\vec{\nabla}\rightarrow\vec{\nabla}+ie\frac{1+\tau_3}{2}\vec{A}$ in
the Lagrangian of eq.(\ref{Eq:8}) with $\vec{A}$ as an external
field.

We have both E1 and PV interactions together, so it is obvious that
the neutron-deuteron system can change from the initial channel
$^2S_{\frac{1}{2}}$ or $^4S_{\frac{3}{2}}$ to the final channel
$^2S_{\frac{1}{2}}$ or $^4S_{\frac{3}{2}}$  at the thermal energy.
Thus, there are four possible transitions,
\begin{eqnarray}\label{Eq:020}
  ^2S_{\frac{1}{2}}\rightarrow\,^2S_{\frac{1}{2}},\quad\quad
  ^2S_{\frac{1}{2}}\rightarrow\,^4S_{\frac{3}{2}},\nonumber\\
  ^4S_{\frac{3}{2}}\rightarrow\,^2S_{\frac{1}{2}},\quad\quad
  ^4S_{\frac{3}{2}}\rightarrow\,^4S_{\frac{3}{2}}\,,
\end{eqnarray}
where the first and third transitions can make $^3$$H$ and $^3$$He$.
So, we have two E1 transitions for the PV neutron radiative capture
by deuteron, corresponding to the $X$ partial wave of the incoming
neutron-deuteron system ($X=^2$$S_{\frac{1}{2}}$ and
$X=^4$$S_{\frac{3}{2}}$).

The spin structure of the E1 matrix element of PV $nd\rightarrow$
$^3H\gamma$ process can be written as
\begin{eqnarray}\label{Eq:00007}
  i\big(t^\dag\sigma_aN\big)\big(\vec{\varepsilon}_d\times\vec{\varepsilon}^\ast_\gamma\big)_a\:,\qquad \big(t^\dag
  N\big)\big(\vec{\varepsilon}_d\cdot\vec{\varepsilon}^\ast_\gamma\big)\,.\;\;\;
\end{eqnarray}

We can obtain for both incoming doublet ($X=^2$$S_{\frac{1}{2}}$)
and quartet ($X=^4$$S_{\frac{3}{2}}$) channels
\begin{eqnarray}\label{Eq:00008}
  \mathcal{W}^{PV}(^2S_{\frac{1}{2}})\big[\;t^\dag\big(\vec{\varepsilon}_d\cdot\vec{\varepsilon}^\ast_\gamma+i\vec{\sigma}\cdot\vec{\varepsilon}_d\times\vec{\varepsilon}^\ast_\gamma\big)N\big]\,,
\nonumber\\
\nonumber\mathcal{W}^{PV}(^4S_{\frac{3}{2}})\big[\;t^\dag\big(\vec{\varepsilon}_d\cdot\vec{\varepsilon}^\ast_\gamma+i\vec{\sigma}\cdot\vec{\varepsilon}_d\times\vec{\varepsilon}^\ast_\gamma\big)N\big]\,.
\end{eqnarray}
where $\mathcal{W}^{PV}(X)$ is the amplitude of the E1 transition
for the incoming $X$ partial wave. In this section, we are going to
calculate $\mathcal{W}^{PV}(X)$ from all possible Feynman diagrams
introduced in figs.\ref{Fig3} and \ref{Fig4} for the PV
$nd\rightarrow$ $^3H\gamma$ process.

If we want to use the PV $Nd$ scattering amplitude for the
calculation of some diagrams in the set (I) and also the set (II),
we should make clear the possible incoming and outgoing partial
waves for PV $Nd$ scattering amplitude.

For the $\mathcal{A}_1$, $\mathcal{A}_2$, $\mathcal{B}$ and
$\mathcal{C}$ diagrams in fig.\ref{Fig4}, we can schematically
represent the PV and E1 transitions as
\begin{eqnarray}\label{Eq:025}
X\;\longrightarrow\!\!\!\!\!\!\!\!\!^{PV}\quad Y
\quad\longrightarrow\!\!\!\!\!\!\!\!\!^{E1}\quad ^2S_{\frac{1}{2}}\,
(triton),
\end{eqnarray}
where above scheme shows the incoming and outgoing partial wave of
the PV $Nd$ scattering are $X$ and $Y$, respectively. At the thermal
energy, $X$ can be $^2$$S_{\frac{1}{2}}$ and $^4$$S_{\frac{3}{2}}$,
as incoming doublet and quartet channels, and $Y$ also can be
$^2$$P_{\frac{1}{2}}$, $^2$$P_{\frac{3}{2}}$, $^4$$P_{\frac{1}{2}}$
or $^4$$P_{\frac{3}{2}}$ (see eq.(\ref{Eq:35})). Furthermore,
because the electric interaction does not change the relative spin
of the particles, all E1 transitions which can create the triton are
\begin{eqnarray}\label{Eq:026}
^2P_{\frac{1}{2}}\longrightarrow\!\!\!\!\!\!\!\!\!^{E1}\quad
^2S_{\frac{1}{2}}\,,\qquad
^2P_{\frac{3}{2}}\longrightarrow\!\!\!\!\!\!\!\!\!^{E1}\quad
^2S_{\frac{1}{2}}\,.
\end{eqnarray}
So, from eqs.(\ref{Eq:35}), (\ref{Eq:025}) and (\ref{Eq:026}), we
finally have only two possible transitions
\begin{eqnarray}\label{Eq:027}
^2S_{\frac{1}{2}}\longrightarrow\!\!\!\!\!\!\!\!\!^{PV}\quad
^2P_{\frac{1}{2}}\longrightarrow\!\!\!\!\!\!\!\!\!^{E1}\quad
^2S_{\frac{1}{2}}\,,\nonumber\\
^4S_{\frac{3}{2}}\longrightarrow\!\!\!\!\!\!\!\!\!^{PV}\quad
^2P_{\frac{3}{2}}\longrightarrow\!\!\!\!\!\!\!\!\!^{E1}\quad
^2S_{\frac{1}{2}}\,,\,
\end{eqnarray}
from the initial S-wave state to triton for the $\mathcal{A}_1$,
$\mathcal{A}_2$, $\mathcal{B}$ and $\mathcal{C}$ diagrams which are
shown in fig.\ref{Fig4}. Hence,
$t^{LO,PV}(^2$$S_{\frac{1}{2}}\rightarrow ^2$$P_{\frac{1}{2}})$ and
$t^{LO,PV}(^4$$S_{\frac{3}{2}}\rightarrow ^2$$P_{\frac{3}{2}})$
amplitudes are needed for calculating the contribution of these
diagrams.

A similar arguments can be expressed for the $E_j$ ($j=1,2,3$), $F$
and $H$ diagrams in the set (I). Consequently, for these diagrams,
one can see that there are two possible transitions which are
presented schematically as
\begin{eqnarray}\label{Eq:027}
^2S_{\frac{1}{2}}\longrightarrow\!\!\!\!\!\!\!\!\!^{E1}\quad
^2P_{\frac{1}{2}}\longrightarrow\!\!\!\!\!\!\!\!\!^{PV}\quad
^2S_{\frac{1}{2}}\,,\nonumber\\
^4S_{\frac{3}{2}}\longrightarrow\!\!\!\!\!\!\!\!\!^{E1}\quad
^4P_{\frac{1}{2}}\longrightarrow\!\!\!\!\!\!\!\!\!^{PV}\quad
^2S_{\frac{1}{2}}\,.\,
\end{eqnarray}
Thus, for these diagrams, we require
$\tilde{t}^{LO,PV}(^2$$P_{\frac{1}{2}}\rightarrow
^2$$S_{\frac{1}{2}})$ and
$\tilde{t}^{LO,PV}(^4$$P_{\frac{1}{2}}\rightarrow
^2$$S_{\frac{1}{2}})$ amplitudes to obtain the contribution of
$E_j$, $F$ and $H$ diagrams in fig.\ref{Fig3}. $\tilde{t}^{LO,PV}$
denotes the PV $Nd$ scattering amplitude corresponding to the
diagrams $a$, $c$ and time-reversed contribution of $a$, $b$ shown
in fig.\ref{Fig2}.

Now, one can be able to begin  the calculation of the sets (I) and
(II) diagrams. Initially, we describe the set (I) (fig.\ref{Fig3})
and then proceed to explain the set (II) (fig.\ref{Fig4}). Let us
consider the PV LECs in eq.(\ref{Eq:8}) and rename them conveniently
from up to down by $g_1$ to $g_5$, respectively; for example, $g_2$
is $g^{(^1S_0-^3P_0)}_{\Delta I=0}$.

The amplitude of all diagrams in fig.\ref{Fig3}, for the incoming X
channel, can be written as
\begin{eqnarray}\label{Eq:0021}
W^{PV}_I(X;E_i,E_f,k)=S(X;E_i,E_f,k)\qquad\qquad\qquad\quad\qquad\qquad\qquad\qquad\qquad
 \nonumber\\ -\frac{1}{2\pi^2}\int
q^2\:dq\:S(X;E_i,E_f,q)\:
\mathcal{D}^{LO}(E_i-\frac{q^2}{2m_N},q)\:t^{LO,PC}(X;E_i,k,q)
,\quad
\end{eqnarray}
where subscript "$I$" in $W^{PV}_I$ represents the contribution of
the set (I) diagrams and $X$ is the partial wave of the incoming
channel (doublet or quartet). We assume
$E_i=E=\frac{3}{4}\frac{k^2}{m_N}-\frac{\gamma^2_t}{m_N}$ and
$E_f=-B_t$.

In the above equation, $S(X;E_i,E_f,k)$ kernel is given by
\begin{eqnarray}\label{Eq:00021}
S(X;E_i,E_f,k)=S_1(X;E_i,E_f,k)+S_2(X;E_i,E_f,k)+S_3(X;E_i,E_f,k),
\end{eqnarray}
where $S_i$ ($i=1,2,3$) is the contribution of all diagrams in the
$i$-th line of fig.\ref{Fig3}. The results of the $S_1$ and $S_2$
kernels for two incoming doublet and quartet channels are presented
in \ref{Appendix B}, however we introduce the $S_3$ kernel in the
following. We note that the groups of the diagrams represented by
$C$, $D$ and $E$ in fig.\ref{Fig3} have two contributions
corresponding to the poles in the nucleon propagators before and
after the photon creation.

The contribution of all diagrams in the line 3 of fig.\ref{Fig3},
for the incoming $X$ channel, in the cluster-configuration space is
given by
\begin{eqnarray}\label{Eq:0000}
S_3(X;E_i,E_f,k)=
 {t^{^3H}}^{\dag}(k)\;
\mathcal{D}^{LO}(E_f-\frac{k^2}{2m_N},k)\:O(X;E_i,E_f,k)+\quad
\nonumber\\ -\frac{1}{2\pi^2}\int q'^2\:dq'
{t^{^3H}}^{\dag}(q')\;\mathcal{D}^{LO}(E_f-\frac{q'^2}{2m_N},q')\times\qquad
\nonumber
\\ \times \tilde{t}^{LO,PV}(Y\rightarrow
^2S_{\frac{1}{2}};E_f,k,q') \qquad\qquad\quad\qquad\nonumber\\
\times
\mathcal{D}^{LO}(E_f-\frac{k^2}{2m_N},k)\;\tilde{O}(X\rightarrow
Y;E_i,E_f,k),\;\;\;
\end{eqnarray}
where, as previously explained in eq.\ref{Eq:027}, $Y$ partial wave
should be $^2P_{\frac{1}{2}}$ and $^4P_{\frac{1}{2}}$ for
$X=$$^2S_{\frac{1}{2}}$ and $^4S_{\frac{3}{2}}$, respectively. So,
we evaluate for the corresponding partial waves the following $O$
matrices,
\begin{eqnarray}\label{Eq:023}
  O(^2S_{\frac{1}{2}};E_i,E_f,q)=\qquad\qquad\qquad\qquad\qquad\qquad\qquad\qquad\qquad\qquad\qquad\qquad\qquad\quad\:\: \nonumber\\
-\frac{eym_N}{18\sqrt{2}\pi}\Bigg\{3\Bigg[\sqrt{\frac{3q^2}{4}-m_NE_f}\left(
\begin{array}{cc}
 -2g_5 & \tau_3g_2 \\
 \tau_3g_1 & 0 \\
\end{array}
\right)+\sqrt{\frac{3q^2}{4}-m_NE_i}
 \left(
\begin{array}{cc}
-2g_5 & \tau_3g_1 \\
 \tau_3g_2 & 0 \\
\end{array}
\right)\Bigg] \nonumber
\\+G(E_i,E_f,q)\left(
\begin{array}{cc}
 4g_5 & -\tau_3(g_1+g_2) \\
 -\tau_3(g_1+g_2) & 0 \\
\end{array}
\right)\Bigg\},\qquad\qquad\qquad\qquad\quad
\nonumber \\
\nonumber \\
O(^4S_{\frac{3}{2}};E_i,E_f,q)=\qquad\qquad\qquad\qquad\qquad\qquad\qquad\qquad\qquad\qquad\qquad\qquad\qquad\quad\:\:\nonumber\\
-\frac{eym_N}{9\sqrt{6}\pi}\Bigg\{3\Bigg[\sqrt{\frac{3q^2}{4}-m_NE_f}\left(
\begin{array}{cc}
 g_5 & 0 \\
 \tau_3g_1 & 0 \\
\end{array}
\right)+\sqrt{\frac{3q^2}{4}-m_NE_i}
 \left(
\begin{array}{cc}
g_5 & 0 \\
 \tau_3g_2 & 0 \\
\end{array}
\right)\Bigg] \nonumber
\\+G(E_i,E_f,q)\left(
\begin{array}{cc}
 -2g_5 & 0 \\
 -\tau_3(g_1+g_2) & 0 \\
\end{array}
\right)\Bigg\}.\qquad\qquad\qquad\qquad\qquad\quad
\end{eqnarray}
The calculated $\tilde{O}(X\rightarrow Y)$ matrix in
eq.(\ref{Eq:0000}) is given for the required partial waves by
\begin{eqnarray}\label{Eq:00028}
  \tilde{O}(^2S_{\frac{1}{2}}\rightarrow^2P_{\frac{1}{2}};E_i,E_f,q)=
-\frac{ey^2}{96\pi}\frac{q}{E_f-E_i}\Bigg[\sqrt{\frac{3q^2}{4}-m_NE_i}-\sqrt{\frac{3q^2}{4}-m_NE_f}\Bigg]\left(
\begin{array}{cc}
 1 & 0 \\
 0 & 1 \\
\end{array}
\right),
\nonumber\\
\tilde{O}(^4S_{\frac{3}{2}}\rightarrow^4P_{\frac{1}{2}};E_i,E_f,q)=
\sqrt{\frac{2}{3}}\tilde{O}(^2S_{\frac{1}{2}}\rightarrow^2P_{\frac{1}{2}};E_i,E_f,q)\left(
\begin{array}{cc}
 1 & 0 \\
 0 & 0 \\
\end{array}
\right),\qquad\qquad\qquad\qquad\qquad\!\!\nonumber\\
\tilde{O}(^2P_{\frac{1}{2}}\rightarrow^2S_{\frac{1}{2}};E_i,E_f,q)=
\tilde{O}(^2S_{\frac{1}{2}}\rightarrow^2P_{\frac{1}{2}};E_i,E_f,q),\qquad\quad\;\;\qquad\qquad\qquad\qquad\qquad\qquad
\nonumber\\
  \tilde{O}(^2P_{\frac{3}{2}}\rightarrow^2S_{\frac{1}{2}};E_i,E_f,q)=
  \frac{2}{\sqrt{3}}\tilde{O}(^2S_{\frac{1}{2}}\rightarrow^2P_{\frac{1}{2}};E_i,E_f,q).\qquad\quad\;\,\qquad\qquad\qquad\qquad\qquad\quad\!\!
\end{eqnarray}
We have also presented the results of
$\tilde{O}(^2P_{\frac{1}{2}}\rightarrow^2S_{\frac{1}{2}})$ and
$\tilde{O}(^2P_{\frac{3}{2}}\rightarrow^2S_{\frac{1}{2}})$ in
eq.(\ref{Eq:00028}) because they are required in the following
calculation regarding the diagram $\mathcal{C}$ in fig.\ref{Fig4}.
In eq.(\ref{Eq:023}) the $G$ function is
\begin{eqnarray}\label{Eq:024}
G(E_i,E_f,q)=\frac{1}{m_N(E_f-E_i)}\bigg\{\frac{3q^2}{4}\left[(\frac{3q^2}{4}-m_NE_i)^{1/2}-(\frac{3q^2}{4}-m_NE_f)^{1/2}\right]\:\:\,\nonumber
\\-\left[(\frac{3q^2}{4}-m_NE_i)^{3/2}-(\frac{3q^2}{4}-m_NE_f)^{3/2}\right]\bigg\}.\quad\,
\nonumber\\
\end{eqnarray}

The above results are evaluated by considering the suitable
partial-wave projections for the incoming and outgoing channels. One
can see all projection operators in the cluster-configuration space
which are used in our calculations in the ref.\cite{G-S-S}.

All other possible diagrams in the calculation of PV $nd\rightarrow$
$^3H\gamma$ amplitude, at the thermal energy, are shown in
fig.\ref{Fig4} as the set (II).

We note that the $\mathcal{A}_2$ diagram in fig.\ref{Fig4} has no
contribution in $\mathcal{W}^{PV}$ because the three-body force
introduced in sect.\ref{PC nd scattering sector} is zero for the
P-wave ($L=1$). The contribution of the $\mathcal{A}_1$ and
$\mathcal{B}$ diagrams in fig.\ref{Fig4}, for two incoming doublet
($X=$$^2$$S_{\frac{1}{2}}$) and quartet ($X=$$^4$$S_{\frac{3}{2}}$)
channels is given by
\begin{eqnarray}\label{Eq:0045}
{W^{PV}_{II}}^{\mathcal{A}_1+\mathcal{B}}(X;E_i,E_f,k)=
-\frac{1}{2\pi^2}\int
q^2\:dq\:\big[\mathcal{A}_1(Y;E_i,E_f,q)+\mathcal{B}(Y;E_i,E_f,q)\big]\times\:\:
\nonumber\\
\times
\mathcal{D}^{LO}(E_i-\frac{q^2}{2m_N},q)t^{LO,PV}(X\rightarrow
Y;E_i,k,q),\quad
\end{eqnarray}
where $\mathcal{A}_1$ and $\mathcal{B}$ can be written as
\begin{eqnarray}\label{Eq:00024}
\mathcal{A}_1(Y;E_i,E_f,q)=a_Y\frac{ey^2}{E_f-E_i}\int
\frac{d^3q'}{(2\pi)^3}\frac{1}{q}\:{t^{^3H}}^{\dag}(q')\times\qquad\qquad\qquad\qquad\qquad\;\nonumber\\
\times\bigg[\mathcal{D}^{LO}(E_i-\frac{q'^2}{2m_N},q')\frac{\:\vec{q}\cdot\vec{q'}}{m_NE_i-q^2-q'^2-\vec{q}\cdot\vec{q'}}\qquad\quad\:\:
\nonumber\\
-\mathcal{D}^{LO}(E_f-\frac{q'^2}{2m_N},q')\frac{\:\vec{q}\cdot\vec{q'}}{m_NE_f-q^2-q'^2-\vec{q}\cdot\vec{q'}}\bigg]\qquad\nonumber\\
\times\left(
\begin{array}{cc}
 -3(1+\tau_3)& 9(1+\tau_3) \\
 3(3-\tau_3) & -(3-\tau_3) \\
\end{array}
\right),\qquad\qquad\qquad\qquad\qquad\quad\!\!\!\nonumber\\
\end{eqnarray}
and
\begin{eqnarray}\label{Eq:00024}
\mathcal{B}(Y;E_i,E_f,q)=a_Y\frac{ey^2}{E_f-E_i}\frac{1}{q}\int
\frac{d^3q'}{(2\pi)^3}\:{t^{^3H}}^{\dag}(q')\times\qquad\qquad\qquad\qquad\qquad\;\nonumber\\
\times\mathcal{D}^{LO}(E_f-\frac{q'^2}{2m_N},q')\bigg[\frac{\:\vec{q}\cdot(\vec{q}+\vec{q'})}{m_NE_i-q^2-q'^2-\vec{q}\cdot\vec{q'}}\qquad\qquad\:\:
\nonumber\\
-\frac{\:\vec{q}\cdot(\vec{q}+\vec{q'})}{m_NE_f-q^2-q'^2-\vec{q}\cdot\vec{q'}}\bigg]
\left(
\begin{array}{cc}
 -3(1-\tau_3)& 3(3+\tau_3) \\
 3(3+\tau_3) & -(3+5\tau_3) \\
\end{array}
\right) .\nonumber\\
\end{eqnarray}
In the above equations, $a_Y$ is $\frac{1}{96}$ for
$Y=\,^2$$P_{\frac{1}{2}}$ and $\frac{1}{32\sqrt{3}}$ for
$Y=\,^2$$P_{\frac{3}{2}}$ partial wave.

The final diagram in the $P_\gamma$ calculation is the $\mathcal{C}$
diagram shown in fig.\ref{Fig4}. After applying the Feynman rules
and the appropriate incoming and outgoing projections, one can see
that the contribution of the $\mathcal{C}$ diagram is given by
\begin{eqnarray}\label{Eq:00026}
{W^{PV}_{II}}^{\;\mathcal{C}}(X;E_i,E_f,k)= -\frac{1}{2\pi^2}\int
q^2\:dq\:{t^{^3H}}^{\dag}(q)\times\qquad\qquad\qquad\qquad\quad\quad\nonumber\\
\times
\mathcal{D}^{LO}(E_f-\frac{q^2}{2m_N},q)\tilde{O}(Y\rightarrow
^2S_{\frac{1}{2}};E_i,E_f,q)\qquad
\nonumber\\
\times\mathcal{D}^{LO}(E_i-\frac{q^2}{2m_N},q)\,t^{LO,PV}(X\rightarrow
Y;E_i,k,q).\quad\,
\end{eqnarray}
In the eqs.(\ref{Eq:0045})-(\ref{Eq:00026}), $Y$ partial wave as
explained in eq.\ref{Eq:027} is $^2$$P_{\frac{1}{2}}$  and
$^2$$P_{\frac{3}{2}}$ for the incoming doublet and quartet channels,
respectively. The calculated results of the
$\tilde{O}(^2P_{\frac{1}{2}}\rightarrow^2$$S_{\frac{1}{2}})$ and
$\tilde{O}(^2P_{\frac{3}{2}}\rightarrow^2$$S_{\frac{1}{2}})$ are
previously introduced in eq.(\ref{Eq:00028}).

We emphasize that the diagrams for the interaction of $H_0$ with a
photon has no contribution because $H_0$ has no derivatives, so it
is not affected by the minimal substitution
$\vec{P}\rightarrow\vec{P}+e\vec{A}$.

Finally, the physical amplitude of all diagrams in the sets (I) and
(II) is given by
\begin{eqnarray}\label{Eq:00025}
\mathcal{W}^{PV}(X;E_i,k)=\Big(W^{PV}_I(X;E_i,k)+W^{PV}_{II}(X;E_i,k)\Big)\left(
                                 \begin{array}{c}
                                   \sqrt{\mathcal{Z}_{LO}} \\
                                   0 \\
                                 \end{array}
                               \right),
\end{eqnarray}
where $W^{PV}_I$ and $W^{PV}_{II}$ are the amplitudes of all
diagrams in the set (I) and (II), respectively. We note that the
value of $\tau_3$ is -1 for $nd\rightarrow$ $^3H\gamma$ as
previously explained in the sect.\ref{PV nd sector}.

\section{Numerical calculation}\label{numerical}
In the calculation of the photon circular polarization, $P_\gamma$,
in $nd\rightarrow$ $^3H\gamma$ process we need to evaluate the total
amplitude of this process. The total amplitude is sum of both PC and
PV contributions that introduced in the previous sections. We
generally classify the numerical computation in three steps.

In the first step, we compute the full-offshell doublet and quartet
PC neutron-deuteron scattering amplitudes and the triton wave
function at LO. The PC neutron-deuteron scattering and the triton
wave function are used in the calculations of the PC and PV
$nd\rightarrow$ $^3H\gamma$ amplitudes. The PC neutron-deuteron
scattering is obtained by solving numerically the Faddeev equations
in eqs.(\ref{Eq:05}) and (\ref{Eq:5}). We solve it by the
Hetherington-Schick method
\cite{Hetherington-Schick,Cahill-Sloan,Aaron-Amado} in a mathematica
code with an arbitrary cutoff momentum $\Lambda$, for detail see
ref.\cite{20 of sadeghi-bayegan}. We also obtain the triton wave
function by solving eq.(\ref{Eq:0005}) as previously explained.

In the second step, we use the off-shell PC neutron-deuteron
scattering amplitude which is computed numerically in the previous
step in order to obtain the PV neutron-deuteron scattering amplitude
of the diagrams in fig.\ref{Fig2}. We have written a new mathematica
code for this step and use the results of the PC neutron-deuteron
scattering amplitude as input data. Note that we choose the same
cutoff momentum $\Lambda$ for the PC neutron-deuteron scattering and
the PV sector calculations.
\begin{figure}
\includegraphics*[width=8cm]{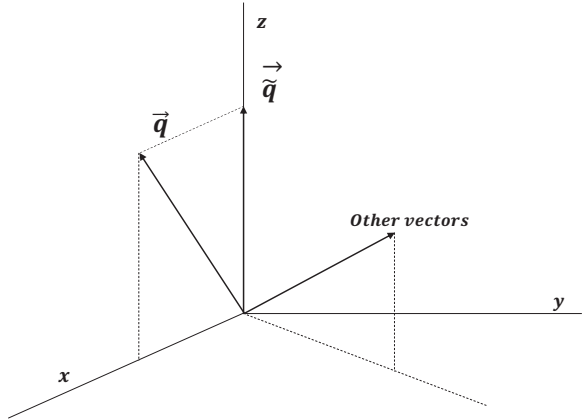} \caption{\label{Fig5}
Coordinate system used in the integration over solid angles.}
\end{figure}

The final step is the calculation of the PC (M1) and PV (E1)
$nd\rightarrow$ $^3H\gamma$ amplitudes. The PC amplitude of
$nd\rightarrow$ $^3H\gamma$ process is obtained by using the PC
neutron-deuteron scattering amplitude and the triton wave function
results.

For the calculation of the PV amplitude we require to evaluate the
set (I) diagrams of PV $nd\rightarrow$ $^3H\gamma$, fig.\ref{Fig3}.
Therefore, we must compute the $S(X;E_i,E_f,q)$ kernel which is
introduced in eq.(\ref{Eq:00021}) (the contribution of the diagrams
in the lines 1, 2 and 3 of set (I)). For the diagrams in the Lines
4, 5 and 6 of the set (I), we add the half-offshell PC
neutron-deuteron scattering amplitude from the left side to the
$S(X;E_i,E_f,q)$ kernel (see eq.(\ref{Eq:0021})).

Finally, the set (II) diagrams in fig.\ref{Fig4} also computed
similar to the set (I) diagrams by using the appropriate
half-offshell PV neutron-deuteron scattering and the triton wave
function data. We solve the integrations numerically by using the
gaussian quadrature weights and the same cutoff $\Lambda$ as before.

We emphasize that for integration over the solid angle, we choose a
coordinate system in which the momentum of outgoing photon is in the
z direction and the $q$ momentum is in the x-z plane (see
fig.\ref{Fig5}).

\section{Results}\label{result}
In this work, we have concentrated on the evaluation of photon
circular polarization, $P_\gamma$, in $nd\rightarrow$ $^3H\gamma$
process and obtaining a new relation, which is useful for extracting
five leading independent LECs.
\begin{figure*}[tb]\centering
\includegraphics*[width=17cm]{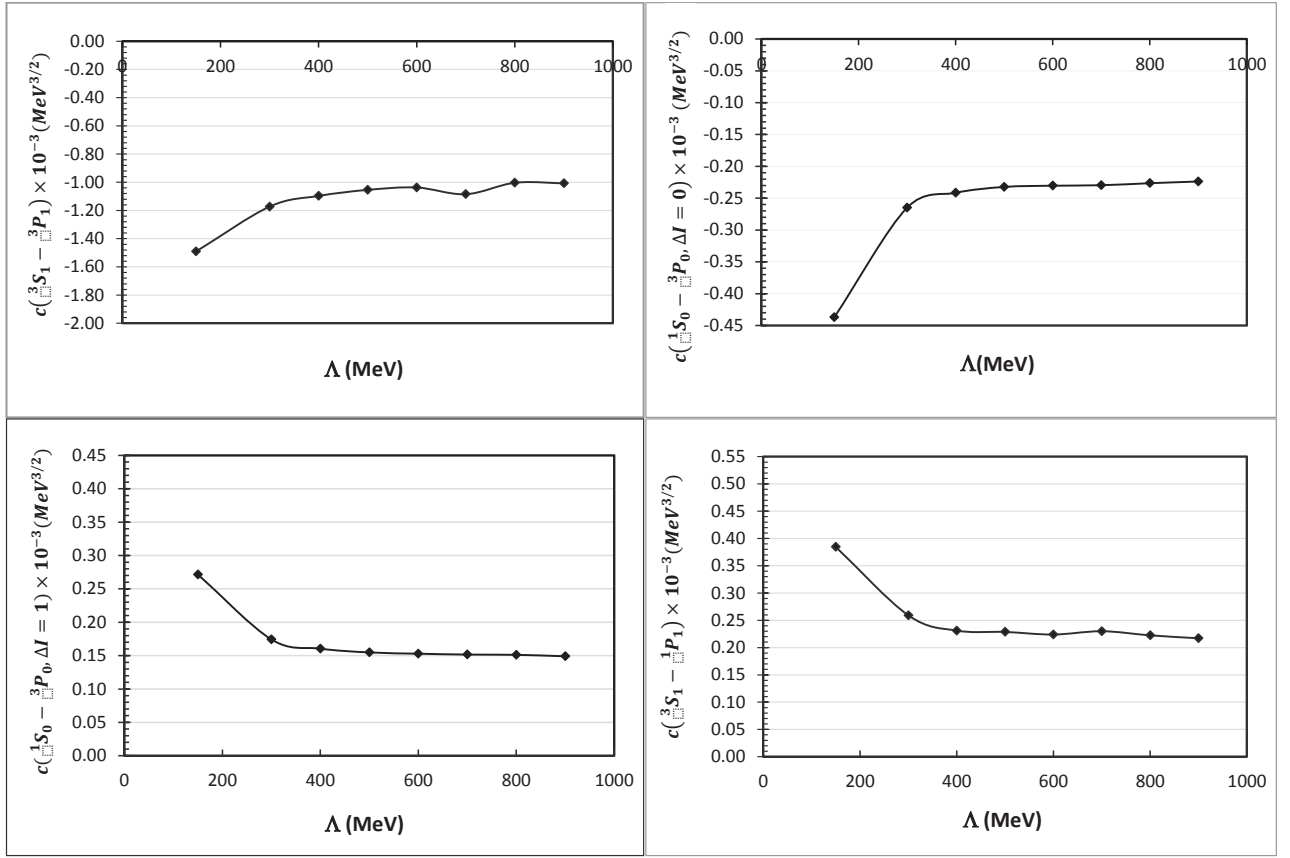} \caption{\label{Fig9}
Variation of the $c(\bar{X}-\bar{Y})$ with cutoff momentum at
thermal energy ($2.5\times10^{-8}\textrm{MeV}$).}
\end{figure*}

The PV polarization of photon is given by
\begin{equation}\label{Eq:32}
P_\gamma=\frac{\sigma_+-\sigma_-}{\sigma_++\sigma_-}\,,
\end{equation}
where $\sigma_+$ and $\sigma_-$ are the cross section for the
photons with right and left helicity, respectively.

We know that the cross sections are calculated with spin summation
on the square amplitude. Therefore one can conclude that the PV
polarization of photon can be written as
\begin{eqnarray}\label{Eq:320}
P_\gamma= 2\frac{\!Re\bigg[{\mathcal{W}^{PC}}^\dagger(^2S_{\frac{1}{2}})\mathcal{W}^{PV}(^2S_{\frac{1}{2}})\!+{\mathcal{W}^{PC}}^\dagger(^4S_{\frac{3}{2}})\mathcal{W}^{PV}(^4S_{\frac{3}{2}})\bigg]}{\big|\mathcal{W}^{PC}(^2S_{\frac{1}{2}})\big|^2+\big|\mathcal{W}^{PC}(^4S_{\frac{3}{2}})\big|^2},\nonumber\\
\end{eqnarray}
where $\mathcal{W}^{PC}(X)$ and $\mathcal{W}^{PV}(X)$ are the PC and
PV amplitudes of $nd\rightarrow$ $^3H\gamma$ process which are
introduced and evaluated in sects.\ref{PC nd radiative capture
sector} and \ref{PV nd radiative capture sector}, respectively.

From eq.(\ref{Eq:320}), one can evaluate the $P_\gamma$  in terms of
the PV LECs. So, for the photon circular polarization, we have
\begin{eqnarray}\label{Eq:032}
P_\gamma=\;\;c(^3S_1-^1P_1)\,g^{(^3S_1-^1P_1)}\qquad\quad\;\;
\nonumber\\+c(^1S_0-^3P_0,\Delta I=0)\,g^{(^1S_0-^3P_0)}_{(\Delta
I=0)}
\nonumber\\
+c(^1S_0-^3P_0,\Delta I=1)\,g^{(^1S_0-^3P_0)}_{(\Delta I=1)}
\nonumber\\+c(^3S_1-^3P_1)\,g^{(^3S_1-^3P_1)},\qquad\quad\,\,
\end{eqnarray}
where $c(\bar{X}-\bar{Y})$ is calculated numerically and shown in
fig.\ref{Fig9}. The $P_\gamma$ does not contain
$g^{(^1S_0-^3P_0)}_{(\Delta I=2)}$ term because the $Nd$ system is
an iso-doublet and PV coupling $g^{(^1S_0-^3P_0)}_{(\Delta I=2)}$
cannot contribute.

According to the table \ref{tab:3} which is shown below, we have
calculated the $Abs[1-\frac{c(\bar{X}-\bar{Y})\; at\;
\Lambda=400\;\textrm{MeV}}{c(\bar{X}-\bar{Y})\; at\;
\Lambda=900\;\textrm{MeV}}]$ for the different incoming and outgoing
partial waves.
\begin{table}[h]\centering
\caption{Results of the cutoff variation of $c(\bar{X}-\bar{Y})$,
for the different incoming and outgoing partial waves, between
$\Lambda=400\;\textrm{MeV}$ and $\Lambda=900\;\textrm{MeV}$.}
\label{tab:3}       
\begin{tabular}{cc}
\hline\hline\noalign{\smallskip} $c(\bar{X}-\bar{Y})$ &
$Abs[1-\frac{c(\bar{X}-\bar{Y})\; at\;
\Lambda=400\;\textrm{MeV}}{c(\bar{X}-\bar{Y})\; at\; \Lambda=900\;\textrm{MeV}}]$  \\
\noalign{\smallskip} \hline\hline\noalign{\smallskip}

$c(^3S_1-^1P_1)$ & $0.0648$  \\

$c(^1S_0-^3P_0,\Delta I=0)$ & $0.0797$ \\

$c(^1S_0-^3P_0,\Delta I=1)$ & $0.0768$ \\

$c(^3S_1-^3P_1)$ & $0.0891$ \\


 \noalign{\smallskip}\hline\hline
\end{tabular}
\end{table}
These results indicate that the cutoff dependence is less than
$0.10$. This small variation can be removed with the consideration
of the higher order corrections.

Therefore, the variation of the $c(\bar{X}-\bar{Y})$ indicates the
ignorable dependence on cutoff at the leading order. So, it can be
concluded that at LO for PV neutron radiative capture by deuteron no
new three-nucleon interaction is needed.

In the numerical calculation for the results shown in \ref{Fig9}, we
used for the nucleon mass $m_N=938.918$ MeV, deuteron binding
momentum $\gamma_t=45.7025$ MeV, effective range of deuteron
$\rho_{d}=1.764$ fm, effective range of $NN$ singlet channel
$r_{0}=2.73$ fm, scattering length in singlet channel of $-23.714$
fm and a triton binding energy $B_t=8.48$ MeV.

It is obvious that the PV coupling constants are not predicted by
the EFT but can be estimated on dimensional grounds. We expect that
the magnitude of the PV coupling to be of the order \cite{G-S-S}
\begin{equation}\label{Eq:033}
\big|g^{(\tilde{X}-\tilde{Y})}\big|\sim 10^{-10}
\:\textrm{MeV}^{-\frac{3}{2}}.
\end{equation}
Of course, we note that this result is dimensional, namely
order-of-magnitude estimation and may be off by factors of 10 or
more. With this estimated value for the PV coupling constants and
order of magnitude of the results of $c(\bar{X}-\bar{Y})$ in
fig.\ref{Fig9}, the value of $P_\gamma$ in $nd\rightarrow$
$^3H\gamma$ is estimated as
\begin{equation}\label{Eq:034}
\big|P_\gamma \big|\sim 10^{-7} .
\end{equation}

Experimentally, the $P_\gamma$ in $nd\rightarrow$ $^3H\gamma$ is not
reported up to now at the thermal energy. Therefore we can compare
our estimated result with a theoretical methods such as DDH model.
The calculated values of $P_\gamma$ in $nd\rightarrow$ $^3H\gamma$
based on DDH model are given by $P_\gamma=-1.39\times10^{-6}$ and
$P_\gamma=-1.14\times10^{-6}$ for two super-soft-core (SSC) and
Reid-soft-core (RSC) potentials, respectively
\cite{desplanques-benayoun}. Recently, the value of $P_\gamma$ in
$nd\rightarrow$ $^3H\gamma$ is calculated with the DDH-best
parameter values and 4-parameter fits for the different potential
models in ref.\cite{Song-L-G}. The order of magnitude of the results
of $P_\gamma$ in the ref.\cite{Song-L-G} is the order of $10^{-7}$.

We can see that our EFT($/\!\!\!\pi$) estimation for the $P_\gamma$
in $nd\rightarrow$ $^3H\gamma$ process at the thermal energy,
eq.(\ref{Eq:034}), agrees with the order of other theoretical
methods.

\section{Conclusion and outlook} \label{conclusion}
In the present EFT($/\!\!\!\pi$) calculation, the $P_\gamma$ is
calculated in order to minimise the uncertainty in determination of
the LECs. We have used the PV two-body interaction in our
calculation. No PV three-body interaction is taken into account at
the leading order.

The value of the coefficient $c(\bar{X}-\bar{Y})$ can provide the
valuable information in order to use in eq.(\ref{Eq:032}). One can
be able to compare eq.(\ref{Eq:032}) with the future experimental
value of $P_\gamma$. The variation of the coefficient
$c(\bar{X}-\bar{Y})$ indicates that the cutoff dependence is
ignorable at the leading order. This small variation can be removed
with the consideration of the higher order corrections.

The photon asymmetry $A_\gamma$ for the neutron-deuteron radiative
capture process is an open problem which we are going to calculate
it in the near future.

The PV proton-deuteron radiative capture is another process which
can help us to identify parity-violating few-body coupling
constants. For this process the coulomb interaction should be
considered together with the PV and EM interactions, simultaneously.

More challenging reactions such as PV four-body scattering will make
the use of EFT more interesting in the future study. This shed light
on the accurate evaluation of weak $dNN$ coupling constants.

\section*{Acknowledgments}
The authors would like to thank Harald W. Grie$\mathcal{B}$hammer
for the valuable $Nd$ scattering mathematica code. This work was
supported by the research council of the University of Tehran.

\setcounter{section}{0} \setcounter{subsection}{0}
\setcounter{equation}{0}
\renewcommand{\theequation}{\Alph{section}.\arabic{equation}}
\renewcommand{\thesection}{Appendix \Alph{section}}
\renewcommand{\thesubsection}{Appendix \Alph{section}.\arabic{subsection}}
\section{The normalization condition for the triton wave
function}\label{Appendix A}

In order to normalize the triton wave function, one can be able to
follow the derivation of the normalization condition for
Bethe-Salpeter equation in \cite{adam,book for normalization}. We
replace the normalization condition of the relativistic two-body BS
vertex function $|\Gamma>$ with the non-relativistic one which is
suitable for the $Nd$ scattering leading to the formation of triton.

One can start from the homogenous equation
\begin{eqnarray}\label{Eq:50}
  |\Gamma>=-VG_{BS}|\Gamma>,
\end{eqnarray}
with the normalization condition introduced by Adam \textit{et al.},
\cite{adam} as
\begin{eqnarray}\label{Eq:51}
  1=-<\Gamma|G_{BS}\frac{\partial}{\partial
  P^2}(VG_{BS})\bigg|_{P^2=M^2}|\Gamma>,
\end{eqnarray}
where $V$ denotes the interaction kernel and $G_{BS}$ represents the
BS propagator. $P$ and $M$ in eq.(\ref{Eq:51}) are total
four-momentum of the system and the bound state mass, respectively.

Based on eq.(\ref{Eq:0005}), the non-relativistic expression similar
to eq.(\ref{Eq:51}) can be written as
\begin{eqnarray}\label{Eq:052}
  1=-\int\frac{d^3q\:dq_0}{(2\pi)^4}\int\frac{d^3q'\:dq'_0}{(2\pi)^4}\qquad\qquad\qquad\qquad\qquad\qquad\:\:
  \nonumber \\ \times (\begin{array}{cc}
                                                            1 & \,0
                                                          \end{array})\big(t^{^3H}(q)\big)^T \mathcal{D}^{LO}(-B_t+q_0,q)\frac{i}{-q_0-\frac{q^2}{2m_N}+i\varepsilon}
                                                          \nonumber \\
                                                          \times \frac{\partial}{\partial
                                                          E}\bigg[V(E,q,q')\mathcal{D}^{LO}(E+q'_0,q')\qquad\qquad\qquad\qquad\,
                                                          \nonumber \\ \times \frac{i}{-q'_0-\frac{q'^2}{2m_N}+i\varepsilon}\bigg]\Bigg|_{E=-B_t} \,t^{^3H}(q')\bigg(\begin{array}{c}
                                                                                                                                  1 \\
                                                                                                                                  0
                                                                                                                                \end{array}
                                                          \bigg).\nonumber
                                                          \\
\end{eqnarray}
where, after taking into account the energy and solid angle
integrations, we have
\begin{eqnarray}\label{Eq:52}
  1=-\int\frac{q^2\,dq}{2\pi^2}\int\frac{q'^2\,dq'}{2\pi^2}\qquad\qquad\qquad\qquad\qquad\qquad\qquad\qquad
  \nonumber \\ \times (\begin{array}{cc}
                                                            1 & \,0
                                                          \end{array})\big(t^{^3H}(q)\big)^T
                                                          \mathcal{D}^{LO}(-B_t-\frac{q^2}{2m_N},q)\qquad\qquad\qquad\quad\;\,
                                                          \nonumber \\
                                                          \times \frac{\partial}{\partial
                                                          E}\Big[V(E,q,q')\mathcal{D}^{LO}(E-\frac{q'^2}{2m_N},q')\Big]\bigg|_{E=-B_t}\!\!\!t^{^3H}(q')\bigg(\begin{array}{c}
                                                                                                                                  1 \\
                                                                                                                                  0
                                                                                                                                \end{array}
                                                          \bigg),\nonumber
                                                          \\
\end{eqnarray}
where $t^{3H}(q)$ has been introduced in the text. we Note that
$\bigg(\begin{array}{c}
    1 \\
    0
  \end{array}
\bigg)$ vector in eq.(\ref{Eq:52}) projects $t^{3H}(q)$ into
incoming dibaryon in the deuteron case. Similarly, for the incoming
dibaryon in the singlet case, $\bigg(\begin{array}{c}
    1 \\
    0
  \end{array}
\bigg)$ and $(\begin{array}{cc}
          1 & \,0
        \end{array})
$ should be replace by $\bigg(\begin{array}{c}
    0 \\
    1
  \end{array}
\bigg)$ and $(\begin{array}{cc}
          0 & \,1
        \end{array})$ in eq.(\ref{Eq:52}), respectively. By comparing eq.(\ref{Eq:0005}) with (\ref{Eq:50}), one can be able to see
\begin{eqnarray}\label{Eq:53}
  V(E,q,q')=2\pi\left[\mathcal{K}^{PC}_{(0)}(E,q,q')\bigg(\begin{array}{cc}
                                                                                                                           1 & -3 \\
                                                                                                                           -3 & 1 \\
                                                                                                                         \end{array}
                                                                                                                       \right)+\mathcal{H}(E,\Lambda)\left(
                                                                                                                                               \begin{array}{cc}
                                                                                                                                                 1 & -1 \\
                                                                                                                                                 -1 & 1 \\
                                                                                                                                               \end{array}
                                                                                                                                             \bigg)\right].
                                                                                                                                             \nonumber
                                                                                                                                             \\
\end{eqnarray}

\setcounter{equation}{0}
\section {The calculation of $S_1$ and $S_2$ kernels of the set (I) diagrams}\label{Appendix B}
We introduce the contribution of all diagrams in the first and
second lines of fig.\ref{Fig3} as $S_{12}$ which is the sum of $S_1$
and $S_2$,
\begin{eqnarray}\label{Eq:0059}
S_{12}(X;E_i,E_f,q)=S_1(X;E_i,E_f,q)+S_2(X;E_i,E_f,q).
\end{eqnarray}
In eq.(\ref{Eq:0059}), $X$ and $q$ are the incoming partial wave and
momentum, respectively. $E_i$ and $E_f$ are the same as introduced
in the text. The $S_{12}$ can be written in the
cluster-configuration space, after applying the energy integration,
as
\begin{eqnarray}\label{Eq:59}
  S_{12}(X;E_i,E_f,q)= -\int q'^2\,dq'\,{t^{3H}}^\dag(q')\bigg\{\mathcal{D}^{
  LO}(E_f-\frac{q'^2}{2m_N},q')\qquad\qquad\qquad
  \nonumber\\ \Big\{\;\;\Big[\mathcal{S}^{A^{up}}+\mathcal{S}^{A^{down}}+\mathcal{S}^{B^{up}}+\mathcal{S}^{B^{down}}+\mathcal{S}^D\Big]\qquad\quad
  \nonumber\\ +\frac{1}{4\pi^4}\int\,p^2\,dp\,\tilde{t}^{LO,PV}(Y\rightarrow
  ^2\!\!S_{\frac{1}{2}};E_f,p
  ,q')\times\qquad
  \nonumber\\ \times\mathcal{D}^{
  LO}(E_f-\frac{p^2}{2m_N},p)\Big[\bar{\mathcal{S}}^{E}+\bar{\mathcal{S}}^{F}\Big]\;\;\Big\}
  \nonumber\\-\mathcal{S}^{C^{up}}-\mathcal{S}^{C^{down}}\bigg\},\qquad\qquad\qquad\qquad\qquad\qquad\quad\;\;\nonumber\\
 \end{eqnarray}
where $\mathcal{S}^{r}$ ($r=A^{up}, A^{down}, B^{up}, B^{down},
C^{up}, C^{down}$, $D$) is a matrix function of $(X;E_i,E_f,q,q')$
and $\bar{\mathcal{S}}^{r'}$ ($r'=E$, $F$) is a matrix function of
$(X\rightarrow Y;E_i,E_f,q,p)$.

We present the equations of all $\mathcal{S}^r$ and
$\bar{\mathcal{S}}^{r'}$ in the following for both incoming doublet
and quartet partial waves.

\subsection {The doublet channel ($X=^2$$S_{\frac{1}{2}}$)}\label{Appendix B.1}
In this section of Appendix B, we introduce the relations of all
$\mathcal{S}^r$ and $\bar{\mathcal{S}}^{r'}$ matrix functions for
the incoming doublet channel. In the following, we have replaced
$g^{(^3S_1-^1P_1)}$, $g^{(^1S_0-^3P_0)}_{\Delta I=0}$,
$g^{(^1S_0-^3P_0)}_{\Delta I=1}$, $g^{(^1S_0-^3P_0)}_{\Delta I=2}$
and $g^{(^3S_1-^3P_1)}$ by $g_1$, $g_2$, $g_3$, $g_4$ and $g_5$,
respectively, as in the text.

  After applying the adequate projections for
$X=^2$$S_{\frac{1}{2}}$ channel, in the cluster-configuration space,
the $\mathcal{S}^{A^{up}}$ and $\mathcal{S}^{B^{up}}$ are given by
\begin{eqnarray}\label{Eq:60}
  \mathcal{S}^{A^{up}}(^2S_{\frac{1}{2}};E_i,E_f,q,q')=\frac{ey}{24\sqrt{2}}\frac{1}{E_f-E_i}\!\int\!\frac{d\Omega_q}{(2\pi)^3}\!\int\!\frac{d\Omega_{q'}}{4\pi}\qquad\qquad\qquad
  \nonumber\\ \times
\Bigg[\!\frac{(2\vec{q'}+\vec{q})\cdot(\vec{q}+\vec{q'})}{m_N\,E_i-q^2-q'^2-\vec{q}\cdot\vec{q'}}-\!\frac{(2\vec{q'}+\vec{q})\cdot(\vec{q}+\vec{q'})}{m_N\,E_f-q^2-q'^2-\vec{q}\cdot\vec{q'}}\Bigg]
\nonumber\\
\times\left(
\begin{array}{cc}
 (1-\tau_3)(3g_1-2g_5) & -(3+\tau_3)g_2+2(1+\tau_3)g_3) \\
 (3+\tau_3)(g_1+2g_5) & -(3+5\tau_3)g_2-2(1+\tau_3)g_3) \\
\end{array}
\right), \nonumber\\
 \end{eqnarray}
\begin{eqnarray}\label{Eq:61}
  \mathcal{S}^{B^{up}}(^2S_{\frac{1}{2}};E_i,E_f,q,q')=\frac{eym_N}{12\sqrt{2}}\int\frac{d\Omega_q}{(2\pi)^3}\int\frac{d\Omega_{q'}}{4\pi}\qquad\qquad\qquad\qquad
 \nonumber\\
\times\frac{1}{m_N\,E_f-q^2-q'^2-\vec{q}\cdot\vec{q'}}\left(
\begin{array}{cc}
 3\tau_3g_1+2g_5 & -\tau_3g_2\\
 -\tau_3g_1-6g_5 & 3\tau_3g_2 \\
\end{array}
\right). \nonumber\\
 \end{eqnarray}
We can also obtain the $\mathcal{S}^{A^{down}}$ and
$\mathcal{S}^{B^{down}}$ for the doublet channel by
\begin{eqnarray}\label{Eq:060}
  \mathcal{S}^{A(B)^{down}}(^2S_{\frac{1}{2}};E_i,E_f,q,q')=\bigg[\mathcal{S}^{A(B)^{up}}(^2S_{\frac{1}{2}};E_f,E_i,q',q)\bigg]^\dagger.
 \end{eqnarray}
For all other $\mathcal{S}^r$ and $\bar{\mathcal{S}}^{r'}$ matrices,
in the incoming doublet channel, we have
 \begin{eqnarray}\label{Eq:62}
  \mathcal{S}^{C^{up}}(^2S_{\frac{1}{2}};E_i,E_f,q,q')=\frac{ey}{24\sqrt{2}}\frac{1}{E_f-E_i}\!\int\!\frac{d\Omega_q}{(2\pi)^3}\!\!\int\!\frac{d\Omega_{q'}}{4\pi}\qquad\qquad\qquad\qquad\qquad\quad \nonumber\\
\times\Bigg[\mathcal{D}^{
  LO}(E_i-\frac{q'^2}{2m_N},q')\frac{(2\vec{q}+\vec{q'})\cdot\vec{q'}}{m_N\,E_i-q^2-q'^2-\vec{q}\cdot\vec{q'}}\qquad\quad
  \nonumber\\ -\mathcal{D}^{
  LO}(E_f-\frac{q'^2}{2m_N},q')\frac{(2\vec{q}+\vec{q'})\cdot\vec{q'}}{m_N\,E_f-q^2-q'^2-\vec{q}\cdot\vec{q'}}\Bigg]\quad
\nonumber\\
\times\left(
\begin{array}{cc}
 (1+\tau_3)(3g_1+2g_5) & -(1+\tau_3)(3g_2+2g_3) \\
 (3-\tau_3)(g_1+2g_5) & -(3-\tau_3)g_2-2(1+\tau_3)g_3 \\
\end{array}
\right),
 \end{eqnarray}

  \begin{eqnarray}\label{Eq:63}
  \mathcal{S}^{C^{down}}(^2S_{\frac{1}{2}};E_i,E_f,q,q')=\frac{ey}{24\sqrt{2}}\frac{1}{E_f-E_i}\int\frac{d\Omega_q}{(2\pi)^3}\qquad\qquad\qquad\qquad\qquad\qquad\qquad
  \nonumber\\
\times\int\frac{d\Omega_{q'}}{4\pi}\Bigg[\mathcal{D}^{
  LO}(E_i-\frac{q'^2}{2m_N},q')\frac{(2\vec{q'}+\vec{q})\cdot\vec{q'}}{m_N\,E_i-q^2-q'^2-\vec{q}\cdot\vec{q'}}\quad\!
  \nonumber\\ -\mathcal{D}^{
  LO}(E_f-\frac{q'^2}{2m_N},q')\frac{(2\vec{q'}+\vec{q})\cdot\vec{q'}}{m_N\,E_f-q^2-q'^2-\vec{q}\cdot\vec{q'}}\Bigg]
\nonumber\\ \times\left(
\begin{array}{cc}
 (1\!+\!\tau_3)(3g_1\!+\!2g_5) & (1\!+\!\tau_3)(3g_1\!+\!2g_5) \\
 -(3\!-\!\tau_3)g_2\!+\!2(1\!-\!\tau_3)g_3 & -(3\!-\!\tau_3)g_2\!+\!2(1\!-\!\tau_3)g_3 \\
\end{array}
\right), \nonumber\\
 \end{eqnarray}
    \begin{eqnarray}\label{Eq:65}
  \mathcal{S}^{D}(^2S_{\frac{1}{2}};E_i,E_f,q,q')=\frac{ey^3m_N}{72\sqrt{2}}\frac{1}{E_f-E_i}\int\frac{d^3l}{(2\pi)^3}\int\frac{d\Omega_q}{4\pi}\int\frac{d\Omega_{q'}}{4\pi}\qquad\qquad\qquad \nonumber\\
\times\Bigg\{\Big[\left(
\begin{array}{cc}
 \mathcal{S}_1 & -\mathcal{T} \\
 \mathcal{S}_1 & -\mathcal{T} \\
\end{array}
\right)\big(n_1(E_i,\vec{q},\vec{l},\vec{q'})-n_1(E_f,\vec{q},\vec{l},\vec{q'})\big)\qquad\,
  \nonumber\\
  +\left(
\begin{array}{cc}
 \mathcal{S}_1 & \mathcal{S}_1 \\
 -\mathcal{T} & -\mathcal{T} \\
\end{array}
\right)\big(n_2(E_i,\vec{q},\vec{l},\vec{q'})-n_2(E_f,\vec{q},\vec{l},\vec{q'})\big)\Big]\Bigg\},
\nonumber\\
 \end{eqnarray}

    \begin{eqnarray}\label{Eq:66}
  \bar{\mathcal{S}}^{E}(^2S_{\frac{1}{2}}-^2P_{\frac{1}{2}};E_i,E_f,q,p)=-\frac{ey^4m_N}{48\,p}\frac{1}{E_f-E_i}\int\frac{d^3l}{(2\pi)^3}\int\frac{d\Omega_q}{4\pi}\int\frac{d\Omega_{p}}{4\pi}\quad\qquad \nonumber\\
\times\left(
\begin{array}{cc}
 -1 & 3 \\
 3 & -1 \\
\end{array}
\right)\Big[n_3(E_i,\vec{q},\vec{l},\vec{p})-n_3(E_f,\vec{q},\vec{l},\vec{p})\Big],\quad
\nonumber\\
 \end{eqnarray}
and
 \begin{eqnarray}\label{Eq:67}
  \bar{\mathcal{S}}^{F}(^2S_{\frac{1}{2}}-^2P_{\frac{1}{2}};E_i,E_f,q,p)=\frac{ey^2}{96\,q}\frac{1}{E_f-E_i}\int\frac{d\Omega_q}{(2\pi)^3}\int\!\frac{d\Omega_{p}}{4\pi}\qquad\qquad\qquad\qquad\qquad \nonumber\\
\times\!\Big[\frac{(\vec{q}+\vec{p})\cdot\vec{q}}{m_N\,E_i\!-\!q^2\!-\!p^2\!-\!\vec{q}\cdot\vec{p}}\!-\!\frac{(\vec{q}+\vec{p})\cdot\vec{q}}{m_N\,E_f\!-\!q^2\!-\!p^2\!-\!\vec{q}\cdot\vec{p}}\Big]
\left(
\begin{array}{cc}
 -3(1-\tau_3) & 3(3+\tau_3) \\
 3(3+\tau_3)  & -(3+5\tau_3) \\
\end{array}
\right).
 \end{eqnarray}
In the eqs.(\ref{Eq:65}) and (\ref{Eq:66}) the $n_1$, $n_2$ and
$n_3$ matrix functions are given by
    \begin{eqnarray}\label{Eq:64}
n_1(E,\vec{q},\vec{l},\vec{q'})=\frac{(2\vec{q'}+\vec{l})\cdot\vec{l}}{m_N\,E\!-\!l^2\!-\!q'^2\!-\!\vec{l}\cdot\vec{q'}}\mathcal{D}^{
  LO}(E-\frac{q'^2}{2m_N},q')\qquad\qquad\quad
\nonumber\\
\times\bigg[\frac{1}{m_N\,E\!-\!q^2\!-\!l^2\!-\!\vec{q}\cdot\vec{l}}\left(
\begin{array}{cc}
 -3(1+\tau_3) & 9(1+\tau_3) \\
 3(3-\tau_3) & -(3-\tau_3) \\
\end{array}
\right)\;\;\:\; \nonumber\\ +\frac{2H_0(\Lambda)}{\Lambda^2}\left(
\begin{array}{cc}
 3(1+\tau_3) & -3(1+\tau_3) \\
 -(3-\tau_3) & (3-\tau_3) \\
\end{array}
\right)\bigg],
\nonumber\\
\nonumber\\
n_2(E,\vec{q},\vec{l},\vec{q'})=\frac{(\vec{q'}+2\vec{l})\cdot\vec{l}}{m_N\,E\!-\!l^2\!-\!q'^2\!-\!\vec{l}\cdot\vec{q'}}\mathcal{D}^{
  LO}(E-\frac{q'^2}{2m_N},q')\qquad\qquad\quad
\nonumber\\
\times\bigg[\frac{1}{m_N\,E\!-\!q^2\!-\!l^2\!-\!\vec{q}\cdot\vec{l}}\left(
\begin{array}{cc}
 -3(1+\tau_3) & 9(1+\tau_3) \\
 3(3-\tau_3) & -(3-\tau_3) \\
\end{array}
\right)\;\;\:\; \nonumber\\ +\frac{2H_0(\Lambda)}{\Lambda^2}\left(
\begin{array}{cc}
 3(1+\tau_3) & -3(1+\tau_3) \\
 -(3-\tau_3) & (3-\tau_3) \\
\end{array}
\right)\bigg],
  \nonumber\\
  \nonumber\\
  n_3(E,\vec{q},\vec{l},\vec{p})=\frac{\vec{p}\cdot\vec{l}}{m_N\,E\!-\!l^2\!-\!p^2\!-\!\vec{l}\cdot\vec{p}}\mathcal{D}^{
  LO}(E-\frac{p^2}{2m_N},p)\qquad\qquad\quad\:\:
\nonumber\\
\times\bigg[\frac{1}{m_N\,E\!-\!q^2\!-\!l^2\!-\!\vec{q}\cdot\vec{l}}\left(
\begin{array}{cc}
 -3(1+\tau_3) & 9(1+\tau_3) \\
 3(3-\tau_3) & -(3-\tau_3) \\
\end{array}
\right)\;\;\,\, \nonumber\\ +\frac{2H_0(\Lambda)}{\Lambda^2}\left(
\begin{array}{cc}
 3(1+\tau_3) & -3(1+\tau_3) \\
 -(3-\tau_3) & (3-\tau_3) \\
\end{array}
\right)\bigg],
\nonumber\\
   \end{eqnarray}
where $\mathcal{S}_1$ and $\mathcal{T}$ are introduced previously in
sect.\ref{PV nd sector}.

We should emphasize that the contribution of the $E_3$ diagram in
fig.\ref{Fig3} is zero because of $H_0=0$ at P-wave ($L=1$).

\subsection {The quartet channel ($X=^4$$S_{\frac{3}{2}}$)}\label{Appendix B.2}
Similar to the doublet channel, we can write the following equations
for the quartet channel. These results are obtained in the
cluster-configuration space after applying the adequate incoming and
outgoing partial-wave projections. All calculated $\mathcal{S}^r$
and $\bar{\mathcal{S}}^{r'}$ matrix functions in the quartet channel
are
\begin{eqnarray}\label{Eq:70}
  \mathcal{S}^{A^{up}}(^4S_{\frac{3}{2}};E_i,E_f,q,q')=\frac{ey}{12\sqrt{6}}\frac{1}{E_f-E_i}\!\int\!\frac{d\Omega_q}{(2\pi)^3}\!\int\!\frac{d\Omega_{q'}}{4\pi}\qquad\qquad\qquad\qquad \nonumber\\
\times\!\Bigg[\!\frac{(2\vec{q'}+\vec{q})\cdot(\vec{q}+\vec{q'})}{m_N\,E_i\!-\!q^2\!-\!q'^2\!-\!\vec{q}\cdot\vec{q'}}\!-\!\frac{(2\vec{q'}+\vec{q})\cdot(\vec{q}+\vec{q'})}{m_N\,E_f\!-\!q^2\!-\!q'^2\!-\!\vec{q}\cdot\vec{q'}}\Bigg]
\nonumber\\
\times\left(
\begin{array}{cc}
 (1-\tau_3)(3g_1+g_5) & 0 \\
 (3+\tau_3)(g_1-g_5) & 0 \\
\end{array}
\right),\qquad\qquad\:\:\qquad\qquad\quad\! \nonumber\\
 \end{eqnarray}
 \begin{eqnarray}\label{Eq:070}
  \mathcal{S}^{A^{down}}\!\!(^4S_{\frac{3}{2}};E_i,E_f,q,q')\!=\!\frac{ey}{12\sqrt{6}}\frac{1}{E_f-E_i}\!\int\!\frac{d\Omega_q}{(2\pi)^3}\!\int\!\frac{d\Omega_{q'}}{4\pi}\!\qquad\qquad\qquad\qquad \nonumber\\
\times\!\Bigg[\!\frac{(2\vec{q}+\vec{q'})\cdot(\vec{q}+\vec{q'})}{m_N\,E_i\!-\!q^2\!-\!q'^2\!-\!\vec{q}\cdot\vec{q'}}-\!\frac{(2\vec{q}+\vec{q'})\cdot(\vec{q}+\vec{q'})}{m_N\,E_f\!-\!q^2\!-\!q'^2\!-\!\vec{q}\cdot\vec{q'}}\Bigg]
\nonumber\\
\times\left(
\begin{array}{cc}
 -(1-\tau_3)g_5 & 0 \\
 2(3+\tau_3)g_2+4(1+\tau_3)g_3 & 0 \\
\end{array}
\right), \qquad\qquad\quad\qquad\!\!\!\nonumber\\
 \end{eqnarray}
\begin{eqnarray}\label{Eq:71}
  \mathcal{S}^{B^{up}}(^4S_{\frac{3}{2}};E_i,E_f,q,q')=\frac{eym_N}{6\sqrt{6}}\int\frac{d\Omega_q}{(2\pi)^3}\int\frac{d\Omega_{q'}}{4\pi}\qquad\qquad\qquad\qquad\qquad
 \nonumber\\
\times\frac{1}{m_N\,E_f-q^2-q'^2-\vec{q}\cdot\vec{q'}}\left(
\begin{array}{cc}
 3\tau_3g_1+g_5 & 0\\
 -\tau_3g_1+3g_5 & 0 \\
\end{array}
\right), \nonumber\\
 \end{eqnarray}
 \begin{eqnarray}\label{Eq:071}
  \mathcal{S}^{B^{down}}(^4S_{\frac{3}{2}};E_i,E_f,q,q')=\frac{eym_N}{3\sqrt{6}}\int\frac{d\Omega_q}{(2\pi)^3}\int\frac{d\Omega_{q'}}{4\pi}\qquad\qquad\qquad\qquad\qquad
 \nonumber\\
\times\frac{1}{m_N\,E_f-q^2-q'^2-\vec{q}\cdot\vec{q'}}\left(
\begin{array}{cc}
 g_5 & 0\\
 \tau_3g_2 & 0 \\
\end{array}
\right), \qquad\quad\nonumber\\
 \end{eqnarray}

 \begin{eqnarray}\label{Eq:73}
  \mathcal{S}^{C^{up}}\!\!(^4S_{\frac{3}{2}};E_i,E_f,q,q')\!=\!\frac{ey}{12\sqrt{6}}\frac{1}{E_f-E_i}\!\int\!\frac{d\Omega_q}{(2\pi)^3}\!\int\!\frac{d\Omega_{q'}}{4\pi}
 \qquad\qquad\qquad\qquad \nonumber\\
\times\Bigg[\mathcal{D}^{
  LO}(E_i-\frac{q'^2}{2m_N},q')\frac{(2\vec{q}+\vec{q'})\cdot\vec{q'}}{m_N\,E_i-q^2-q'^2-\vec{q}\cdot\vec{q'}}\:\:
  \nonumber\\ -\mathcal{D}^{
  LO}(E_f-\frac{q'^2}{2m_N},q')\frac{(2\vec{q}+\vec{q'})\cdot\vec{q'}}{m_N\,E_f-q^2-q'^2-\vec{q}\cdot\vec{q'}}\Bigg]
\!\!\!\nonumber\\ \times\left(
\begin{array}{cc}
 (1\!+\!\tau_3)(3g_1\!+\!g_5) &0 \\
 (3\!-\!\tau_3)(g_1\!-\!g_5) & 0 \\
\end{array}
\right),\qquad\qquad\qquad\qquad\quad
 \end{eqnarray}

  \begin{eqnarray}\label{Eq:74}
  \mathcal{S}^{C^{down}}(^4S_{\frac{3}{2}};E_i,E_f,q,q')=\frac{ey}{6\sqrt{6}}\frac{1}{E_f-E_i}\int\frac{d\Omega_q}{(2\pi)^3}\!\int\frac{d\Omega_{q'}}{4\pi}\qquad\qquad\qquad\qquad
  \nonumber\\
\times\Bigg[\mathcal{D}^{
  LO}(E_i-\frac{q'^2}{2m_N},q')\frac{(2\vec{q'}+\vec{q})\cdot\vec{q'}}{m_N\,E_i-q^2-q'^2-\vec{q}\cdot\vec{q'}}\;\;
  \nonumber\\ -\mathcal{D}^{
  LO}(E_f-\frac{q'^2}{2m_N},q')\frac{(2\vec{q'}+\vec{q})\cdot\vec{q'}}{m_N\,E_f-q^2-q'^2-\vec{q}\cdot\vec{q'}}\Bigg]\!\!\!
\nonumber\\ \times\left(
\begin{array}{cc}
 2(1\!+\!\tau_3)g_5 &0 \\
 (3\!-\!\tau_3)g_2\!-\!2(1\!-\!\tau_3)g_3 & 0 \\
\end{array}
\right),\qquad\qquad\qquad\quad
 \end{eqnarray}

 \begin{eqnarray}\label{Eq:77}
  \bar{\mathcal{S}}^{F}(^4S_{\frac{3}{2}}-^4P_{\frac{1}{2}};E_i,E_f,q,p)=\frac{ey^2\sqrt{3}}{24\sqrt{2}\,q}\frac{1}{E_f-E_i}\int\frac{d\Omega_q}{(2\pi)^3} \!\int\!\frac{d\Omega_{p}}{4\pi}\!\qquad\qquad\qquad\qquad\nonumber\\
\!\times\Big[\frac{(\vec{q}+\vec{p})\cdot\vec{q}}{m_N\,E_i\!-\!q^2\!-\!p^2\!-\!\vec{q}\cdot\vec{p}}\!-\!\frac{(\vec{q}+\vec{p})\cdot\vec{q}}{m_N\,E_f\!-\!q^2\!-\!p^2\!-\!\vec{q}\cdot\vec{p}}\Big]
\nonumber\\ \times\left(
\begin{array}{cc}
 (1-\tau_3) & 0 \\
 0 & 0 \\
\end{array}
\right). \qquad\qquad\qquad\qquad\qquad\qquad\quad
 \end{eqnarray}
We introduce two contributions in eqs.(\ref{Eq:62}-\ref{Eq:67}) and
(\ref{Eq:73}-\ref{Eq:77}) for the poles in the nucleon propagators
before and after the photon creation, as mentioned in the text. We
note that the groups of the diagrams represented by $D$ and $E$ in
fig.\ref{Fig3} have no contribution in the incoming quartet channel
because in these diagrams the nucleon which interacts with the
photon is a neutron (for $X=^4$$S_{\frac{3}{2}}$ channel) and a
photon has no E1 interaction with a neutron.

\end{document}